\documentclass[reprint,aps,superscriptaddress,citeautoscript]{revtex4-1}
\usepackage{graphicx}
\usepackage{setspace,transparent}
\usepackage{color}
\usepackage{placeins}
\usepackage{fontenc,ragged2e,wrapfig}
\usepackage{mathtools,amssymb,amsmath,nicefrac,mathrsfs}
\usepackage[font=footnotesize]{caption}
\usepackage{soul}
\usepackage[backref=page]{hyperref}
\setcitestyle{super}

\usepackage[dvipsnames]{xcolor}
\usepackage[utf8]{inputenc}
\usepackage{listings}

\DeclareMathAlphabet{\mathpzc}{OT1}{pzc}{m}{it}
\DeclareCaptionJustification{justified}{\justifying}
\newcommand{\RNum}[1]{\uppercase\expandafter{\romannumeral #1\relax}}

\newsavebox\mysavebox

\begin{document}

\title{Interdependent Superconducting Networks}

\author{\textsc{I.\ Bonamassa*}}
\affiliation{Department of Physics, Bar-Ilan University, 52900 Ramat-Gan, Israel}
\affiliation{Department of Network and Data Science, CEU, Quellenstrasse 51, A-1100 Vienna, Austria}
\email{I.B., B.G.\ \& M.L.\ contributed equally to this work. \\ 
Correspondence should be send to: bonamassai@ceu.edu}
\author{\textsc{B.\ Gross*}}
\affiliation{Department of Physics, Bar-Ilan University, 52900 Ramat-Gan, Israel}
\author{\textsc{M.\ Laav*}}
\affiliation{Department of Physics, Bar-Ilan University, 52900 Ramat-Gan, Israel}
\author{\textsc{I.\ Volotsenko}}
\affiliation{Department of Physics, Bar-Ilan University, 52900 Ramat-Gan, Israel}
\author{\textsc{A.\ Frydman}}
\affiliation{Department of Physics, Bar-Ilan University, 52900 Ramat-Gan, Israel}
\author{\textsc{S.\ Havlin}}
\affiliation{Department of Physics, Bar-Ilan University, 52900 Ramat-Gan, Israel}

\date{\today}

\maketitle
\begin{spacing}{0.91} 

\textbf{\small
Cascades are self-amplifying processes~\cite{motter2017unfolding} triggered by feedback mechanisms that may cause a substantial part of a macroscopic system to change its phase in response of a relatively small local event.
The theoretical background for these phenomena is rich and interdisciplinary~\cite{pocock2012robustness, fornito2015connectomics, hokstad2012risk, helbing2013globally, klosik2017interdependent} \!\!, with interdependent networks~\cite{buldyrev-nature2010, parshani-prl2010} providing a versatile ``two-interactions'' framework to study their multiscale evolution. 
Yet, physics experiments aimed at validating this ever-growing volume of predictions have remained elusive, hitherto hindered by the problem of identifying possible physical mechanisms realizing interdependent couplings. 
Here we develop and study the first experimental realization of an interdependent system as a multilayer network of two disordered superconductors separated by an insulating film.
We show that Joule heating effects~\cite{gurevich1987self} emerging at sufficiently large driving currents act as dependency links between the superconducting layers, igniting overheating cascades via adaptive back and forth electro-thermal feedbacks. 
Through theory and experiments, we unveil a rich phase diagram of mutual resistive transitions and cascading processes that physically realize and generalize interdependent percolation. 
The present work establishes the first physics laboratory bench for the manifestation of the theory of interdependent systems, enabling experimental studies to control and to further develop the multilayer phenomena of complex interdependent materials.} 
\end{spacing}

{\small
Catastrophic events like power-grid outages~\cite{yang2017small, schafer2018dynamically} or regime shifts in urban infrastructures~\cite{rinaldi-ieee2001, little2002controlling, rosato-criticalinf2008} and other ecosystems~\cite{scheffer2009critical, haldane2011systemic, rocha2018cascading} are often the aftermath of cascading processes~\cite{kitsak2010identification, borge2013cascading, morone2015influence} spreading within and across multiple layers. 
Interdependent network theory~\cite{buldyrev-nature2010, bianconi2018multilayer} has granted solid grounds to study these multiscale phenomena, translating the mechanisms fuelling the propagation of avalanches into the interplay between two qualitatively different types of couplings: {\em connectivity} links~\cite{barabasi2016network} \!\!, characterizing the interactions between nodes within layers, and {\em dependency} links, modelling, instead, functional interactions (e.g.\ positive feedback) among nodes between layers\footnote{Therefore, while connectivity links enable physical processes (e.g.\ electric currents) to propagate within layers by hopping from one node to another, dependency links make those processes to influence each other, without though providing a pathway to hop from layer to layer.} \!\!.

Despite many theoretical efforts made in applying this ``two-interactions'' scheme to processes as diverse as percolation~\cite{baxter2012avalanche, bashan-naturephysics2013, radicchi2015percolation} \!\!, dynamics~\cite{nicosia2017collective, danziger2019dynamic} \!\!, transport~\cite{morris2012transport, gross2021interdependent} and elsewhere~\cite{wang2013interdependent} \!\!, developing physics laboratory realizations of interdependent systems has remained for over a decade a fundamental and yet elusive challenge, disabling experimental studies to scrutinize and to further develop the interdisciplinary volume of models and predictions collected so far. 

In this Letter we present the experimental and theoretical characterization of the first physical interdependent material based on a multilayer network composed by two disordered superconductors separated by a thermally-conducting electrical insulator. 
Through a model of thermally-coupled networks of Josephson junctions, we elucidate the mutual percolation processes that underlie the discontinuous onset and fall of global phase coherence observed in our experiments. 
We disclose fundamental and yet overlooked features of interdependent interactions related to their spontaneous emergence, the strength of their action and the suppressive effect they have in the process of functional revival. 
These results establish a laboratory-controlled benchmark that ``breaths life'' in the theory of interdependent systems and enhances our understanding of their complexity beyond modeling. 

\begin{figure*}
	\includegraphics[width=0.85\linewidth]{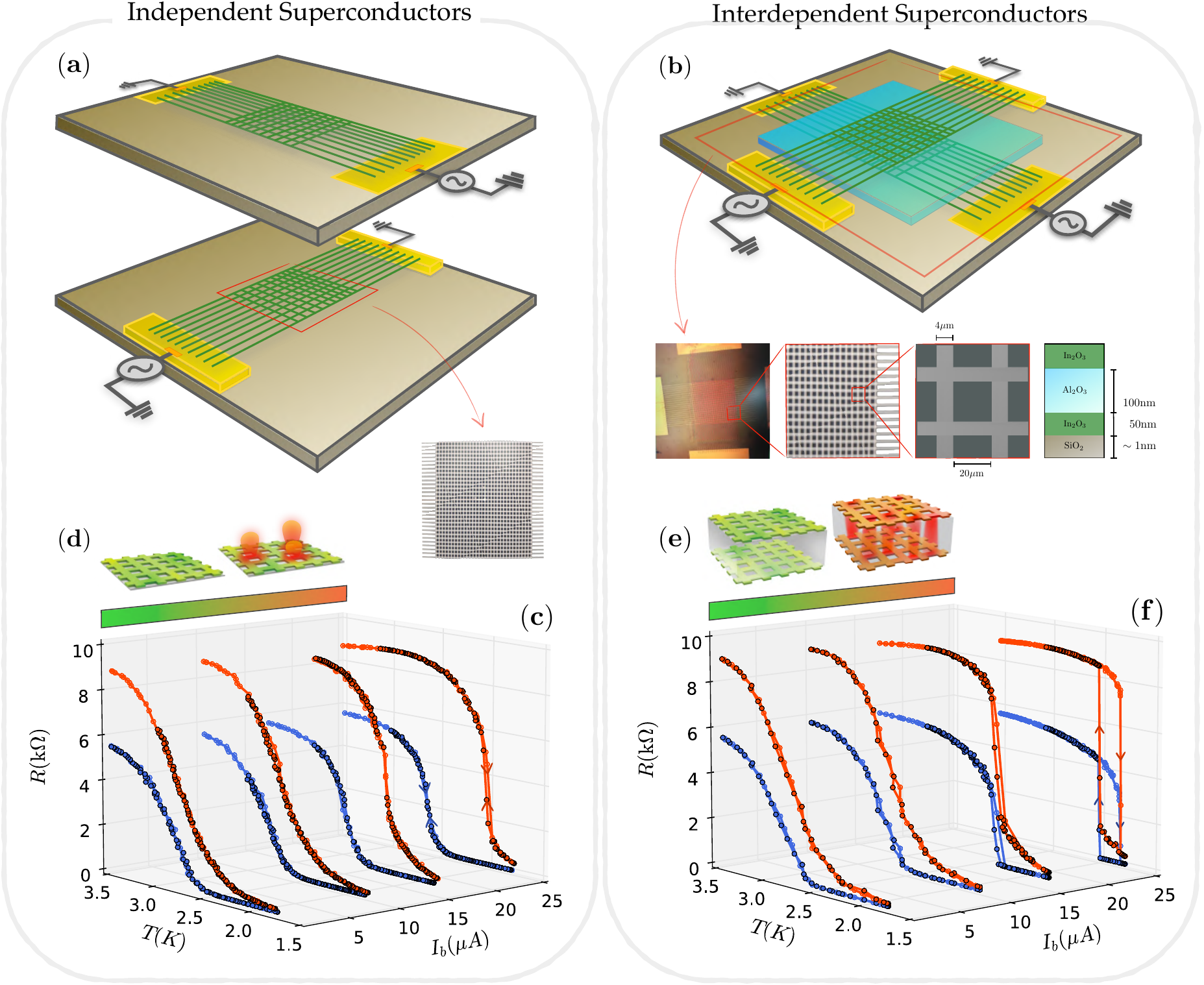}
		\caption{\footnotesize{\textbf{Design and experimental setup of thermally-interdependent superconducting networks.} (Color online) 
		(\textbf{a}) Schematic representation and scanning electron micrograph (zooming inset) of the isolated layers, each involving a $2D$ lattice of disordered superconductors (a:InO films on $\mathrm{SiO}_2$ substrates, see M1 in the Methods for details). The edges of each network are connected to $\mathrm{Au}/\mathrm{Ti}$ contacts.
		(\textbf{b}) Design of the interdependent superconducting material, with bottom and top layers (green grids) separated by a thermally-conducting insulating film ($\mathrm{Al}_2\mathrm{O}_3$). (Inset) Scanning electron micrograph of the interdependent sample and characterization of the layers' physical dimensions. 
		(\textbf{c}) Experimental sheet resistances measured (Methods, M3) in the {\em isolated} top (red) and bottom (blue) layers under identical driving currents at zero magnetic field, for increasing (filled symbols) and decreasing (empty symbols) values of the cryostat temperature $T$ (see the arrows along the curves at $I_b=24\mu\mathrm{A}$). Both layers undergo continuous SN transitions at different bulk critical temperatures. 
		(\textbf{d}) Illustration describing the emergence of local normal-metal (N) hotpots at the SN transition of single layers for large enough driving currents. 
		(\textbf{e}) In the interdependent setup, local hot-spots thermally intertwine the superconducting (SC) states of superposed junctions, physically realizing the cross-layers dependency links. 
		(\textbf{f}) Sheet-resistance measured in the {\em interdependent} superconducting networks for the same set of driving currents displayed in Fig.~\ref{fig:1}\textbf{c}. At large currents (i.e.\ $I_b\gtrsim 15\mu\mathrm{A}$) the layers become thermally-locked in their bulk critical temperatures and undergo unconventional {\em mutual} {\em first}-{\em order} SN transitions. 
		}} \label{fig:1}\vspace*{-0.3cm}
\end{figure*}

\paragraph*{\small \underline{Experimental results}} 
Fig.~\ref{fig:1} shows the schematic design of our multilayer material composed by two disordered superconductors~\cite{saito2016highly} in two configurations: {\em independent} networks (Fig.~\ref{fig:1}\textbf{a}) and {\em thermally-interdependent} networks (Fig.~\ref{fig:1}\textbf{b}), where cross-layer couplings set in through an electrically insulating film with good thermal conductivity ($\mathrm{Al}_2\mathrm{O}_3$). 
Each layer is composed (see Methods, M1) of an e-beam evaporated amorphous indium oxide (a:InO) film, i.e.\ a disordered superconductor characterized by a broad superconducting transition with a bulk critical temperature, $T_c \simeq 3K$, determined by the onset of global phase coherence. \\
\indent 
The experimental results presented in Fig.~\ref{fig:1}\textbf{c},\textbf{f} can be summarized as follows. When measured independently (Methods, M3) and under identical conditions, as illustrated schematically in Fig.~\ref{fig:1}\textbf{a}, each layer undergoes a continuous and broad superconductor-normal (SN) phase transition~\cite{sacepe2011localization, doron2020critical} at some finite bulk critical threshold, $T_c$, whose value depends on the disorder of the sample and on the driving current, $I_b$, flowing through it. 
Since the layers have different levels of disorder, they exhibit different values of $T_c$. The broad SN-transitions become sharper for increasing values of $I_b$ but they always remain continuous and non-hysteretic (Fig.~\ref{fig:1}\textbf{c}). 
On the other hand, when similar sufficiently large $I_b$'s flow simultaneously in both layers, as illustrated in Fig.~\ref{fig:1}\textbf{b}, thermal couplings set in between the networks and their SN transitions become mutually abrupt and hysteretic (Fig.~\ref{fig:1}\textbf{f}). 
Furthermore, we find that the mutual superconducting (SC) to normal-metal (N) transitions are dominated by the high disordered network (having lower $T_c$), while the mutual N-to-SC (NS) jumps are governed instead by the resistive behavior of the low disordered array (having higher $T_c$).

\paragraph*{\small \underline{Two-interactions mechanism}} 
The unconventional discontinuous SC-transitions reported in the experiments can be very well understood within the framework of interdependent networks~\cite{buldyrev-nature2010, parshani-prl2010} \!\!. 
Each independent layer realizes a typical instance of the disorder via its characteristic distribution of critical temperatures $\{T_{{ij}}^c\}_{i,j\leq N}$ and currents $\{I_{ij}^c\}_{i,j\leq N}$, with $N$ being the number of nodes in the $2D$ lattices, which control the SN-activation of single junctions due to the Josephson effect. 
It follows that the bulk critical temperature $T_c$ of these disordered media~\cite{havlin1987diffusion} corresponds to the threshold (at constant $I_b$) where SC-clusters {\em continuously} percolate~\cite{kirkpatrick1973percolation, coniglio1982cluster, skvortsov2005superconductivity} \!\!. 
The experimental independent sheet resistances displayed in Fig.~\ref{fig:1}\textbf{c} confirm the continuous and reversible nature of the SN transition in the isolated SC networks at two independent values of $T_c$ over the range of driving currents tested. 
\begin{figure*}
	\includegraphics[width=0.85\linewidth]{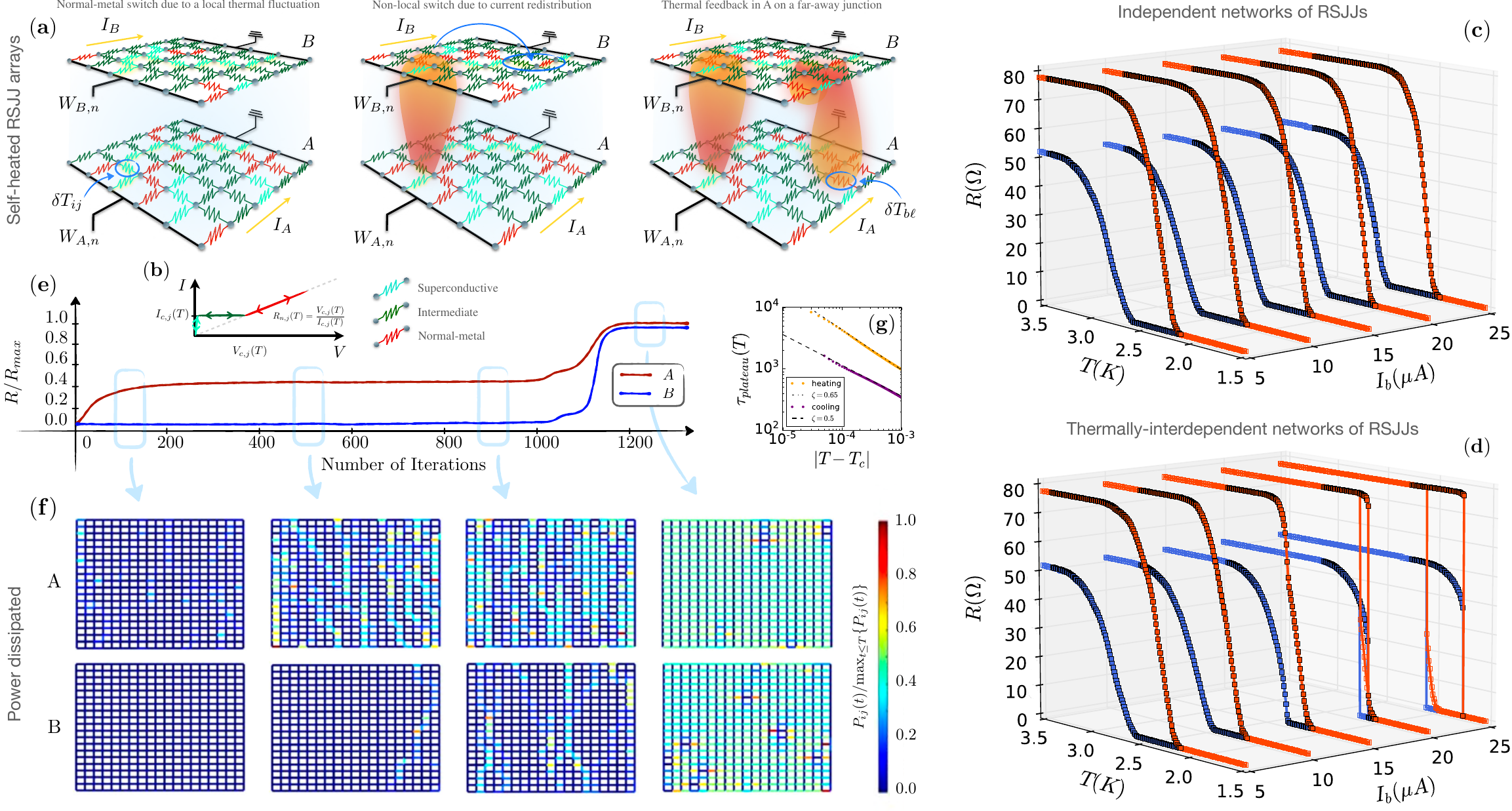}
		\caption{
		\footnotesize {\bf Thermally-interdependent networks of resistively-shunted Josephson junctions.} (Color online) 
		(\textbf{a}) Illustration of the electro-thermal runaway caused by the local hotspots and the redistribution of the currents when transitioning from the mutual SC-phase to the mutual N-phase. A description of the cascading stages in the heating (SC-to-N) and cooling (N-to-SC) processes is given in the text. 
		(\textbf{b}) Josephson $I$-$V$ characteristic adopted to model the switching of single junctions between their electronic states (see Methods, M4).
		(\textbf{c}) Continuous SN-transitions in the global resistances of $2$ thermally-decoupled arrays with $1860$ junctions with mean\vspace*{-0.07cm} critical thresholds $I_{0,A}^c=48\mu\mathrm{A}$ and $T_c^A=2.4\mathrm{K}$ (top layer, red symbols), $I_{0,B}^c=58\mu\mathrm{A}$ and $T_c^B=2.9\mathrm{K}$ (bottom layer, blue symbols), and identical variance\vspace*{-0.05cm} $\sigma_A=\sigma_B=0.1$ due to the equal distribution of disorder in the fabricated samples. We adopt the normal resistive factors (Methods, M4) $\rho_{A}=1.24$ and $\rho_{B}=0.77$, matching the experimental ratio $R_A/R_B\simeq1.61$. 
		(\textbf{d}) Mutual resistive transitions in interdependent RSJJs networks obtained by solving numerically the thermally-coupled Kirchhoff equations (Methods, Eqs.~\eqref{eq:M2},\eqref{eq:M3}) set by the two-interactions interplay between Eq.~\eqref{eq:0} and Eq.~\eqref{eq:1} with $\gamma=3\times10^7\mathrm{K}\mathrm{W}^{-1}$ and $\gamma'=5\times10^5\mathrm{K}\mathrm{W}^{-1}$. The layers' physical parameters are identical to those adopted in (\textbf{c}). 
		(\textbf{e}) Evolution of the marginally stable mutual SC-phase for $I_b=24\mu\mathrm{A}$ slightly above the heating first-order SN threshold, $T_{c,>}=2.08\mathrm{K}$, displaying the emergence of a long-lived plateau with nearly constant resistances. 
		(\textbf{f}) Stroboscopic snapshots of the power dissipated by single junctions during the heating plateau in (\textbf{e}); notice the propagation of local hot-spots (dark red links) between the layers $A$ and $B$ due to the overheating cascades. 
		(\textbf{g}) Metastable lifetime, $\tau(T)$, at $I_b=24\mu\mathrm{A}$ of the heating (orange) and cooling (purple) plateau close to the heating threshold ($T_{c,>}=2.08\mathrm{K}$) and to the cooling one ($T_{c,<}=1.88\mathrm{K}$), respectively. Notice the two scaling exponents hinting at the different growth processes (see the main text for details) that underlie the two jumps of the abrupt transition (see also Fig.~\ref{fig:S4}). 
		}\label{fig:2}\vspace*{-0.3cm}
\end{figure*}
Let us now consider the interdependent scheme (Fig.~\ref{fig:1}\textbf{b}). 
Slightly above their independent $T_c$'s, single layers lack the percolation of SC-paths, hosting, instead, the flow of  dissipative currents. Single junctions switching to the N-state become, therefore, localized hot-spots (Fig.~\ref{fig:1}\textbf{d}) randomly distributed across each array, whose dissipated heat depends on how much current flows through them. 
The $\mathrm{Al}_2\mathrm{O}_3$ medium (Fig.~\ref{fig:1}\textbf{b}, inset) couples thermally the two networks by mediating the hot-spots' heat between the layers while inhibiting the tunneling of electrons. 
In this configuration, physical dependency links spontaneously emerge between the layers in the form of adaptive thermal couplings sustained by Joule dissipation~\cite{danziger-newjphysics2015, bonamassa2021realizing} which thermally intertwine the SC-states of superposed junctions (Fig.~\ref{fig:1}\textbf{e}). 
More concretely, once a junction $a_{ij}$ in layer $A$ switches into metal (Fig.~\ref{fig:2}\textbf{a}), it overheats its superposed junction $b_{ij}$ in layer $B$, thus raising the vulnerability of the latter to exceed its critical temperature. 
This outcome causes a redistribution of the currents in layer $B$ that can activate other junctions, e.g.\ $b_{\ell m}$, as they cross their critical currents, creating more hot-spots that heat back their counterparts, i.e.\ $a_{\ell m}$, in layer $A$. 
This positive electro-thermal runaway, that physically realizes the two-interactions interplay theorized in interdependent percolation~\cite{buldyrev-nature2010,parshani-prl2010} \!\!, ignites avalanches of switching junctions whose {\em non}-{\em local}\,~\footnote{The different realizations of disorder in layers $A$ and $B$ ensure that $a_{ij}$ and $a_{\ell m}$ are statistically independent, thus enabling a {\em non-local} propagation of {\em local} phase perturbations. See Fig.~\ref{fig:2}\textbf{a} for a pictorial representation of the electro-thermal runaway process.} growth across the layers can encompass a large fraction of the system's size, causing the mutual first-order SN-transitions displayed in Fig.~\ref{fig:1}\textbf{f}. 
In particular, heating the networks from low temperatures realizes the propagation of damage created by cascading failures (here, N-states) in interdependent percolation, yielding the mutual fragmentation of the SC-phases in both layers.
On the other hand, when cooling the system from its mutual N-phase, thermal interdependence defers the formation of global phase-coherence to temperatures below the $T_c$'s of the isolated arrays, producing areas of hysteresis. 
In this cooling process, the dissipating hot-spots sustain the mutual N-phase by suppressing the merging of SC-clusters, realizing a mechanism opposite to cascading failure that is analogous to spanning-cluster-avoiding percolation~\cite{cho2013avoiding} \!\!.  
In fact, we provide support based on our theoretical model below (Fig.~\ref{fig:2}\textbf{a}), that compact SC-clusters become dense at low $T$ until they suddenly merge into a giant percolating SC-component (see also Fig.~\ref{fig:S3} and Fig.~\ref{fig:S4} displayed in the Extended Data (ED) section and Supplementary Movies S1, S2), yielding an abrupt onset of global coherence in both arrays that nicely reproduces the mutual NS transitions found in the experiments (Fig.~\ref{fig:1}\textbf{f}). 


\begin{figure*}
	\includegraphics[width=0.85\linewidth]{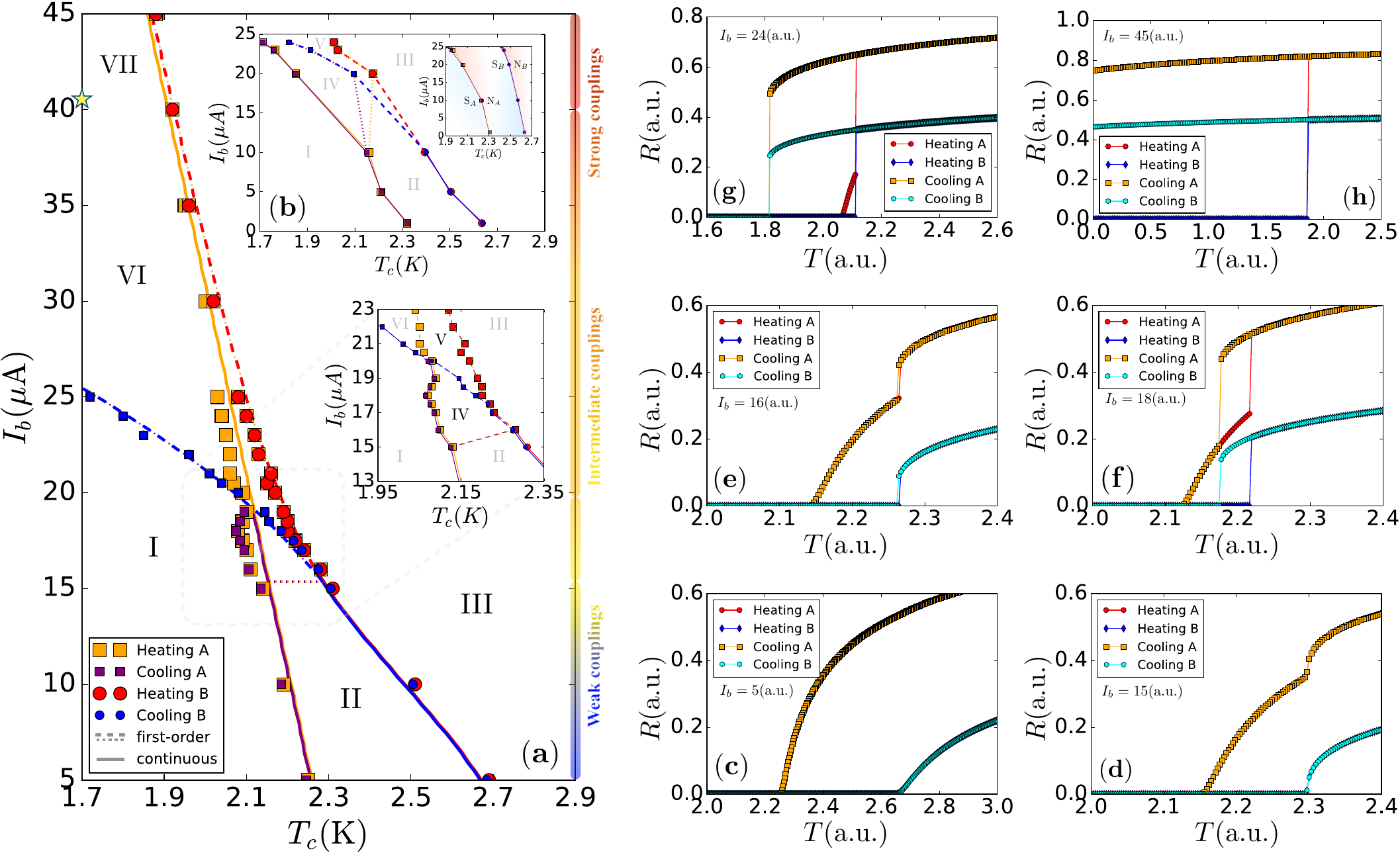}
		\caption{\footnotesize{\textbf{Mutual phase diagram: theory vs.\ experiments}. (Color online) 
		(\textbf{a}) Theoretical thresholds (symbols) characterizing the mutual phase transitions in the globally-interdependent Kirchhoff equations vs.\ the analytical thresholds (curves) obtained via the interdependent HN-resistances, Eq.~\eqref{eq:4}. To solve the latter, the parameters $R_0$, $\beta$, $\omega$, $T_{c,0}$ and $\gamma$ (measured in arbitrary units, $(a.u.)$) have been extracted by best-fitting $R_{HN}$ to the numerical resistive curves of the two layers, taken independently, yielding:\vspace*{-0.06cm} $\beta_A=0.40$, $\beta_B=0.65$, $T_{c,0}^A=2.30$, $T_{c,0}^B=2.75$, $\omega=10^{-2}$ and $\gamma=4\times10^{-3}$; we set $R_0^A=1$ and $R_0^B=0.65$, matching the experimental ratio (caption, Fig.~\ref{fig:2}\textbf{c}).\vspace*{-0.05cm} (Inset) Mutual stability: (\RNum{1}) mutual SC-phase, i.e.\ both $A$ and $B$ are stable superconductors; (\RNum{2}) and (\RNum{4}) $A$ is a stable metal and $B$ is a stable superconductor; (\RNum{3}) mutual N-phase, i.e.\ both $A$ and $B$ are stable metals. Mutual metastability: (\RNum{5}) $A$-partial-N and $B$-SC with the mutual N-phase; (\RNum{6}) and (\RNum{7}) mutual N-phase with mutual SC-phase. Phases (\RNum{4}) and (\RNum{7}) differ from, respectively, phases (\RNum{2}) and (\RNum{6}) in their mutual transitions [see plots (\textbf{c})--(\textbf{h})]. 
		(\textbf{b}) Experimental phase diagram extracted from the resistive curves displayed in Fig.~\ref{fig:1}\textbf{f}. (Inset) Decoupled thresholds describing the continuous SN-transitions in Fig.~\ref{fig:1}\textbf{c}. 
		(\textbf{c})--(\textbf{h}) Analytical mutual transitions at increasing interdependence strengths. Weak couplings, (\textbf{c}) and (\textbf{d}): $A$ and $B$ undergo nearly-independent continuous SN transitions until a cusp forms in layer $B$. Moderate couplings, (\textbf{e}) and (\textbf{f}): $A$ undergoes a two-steps transition, with a continuous step from a SC-phase to a partial N-phase and a first-order jump to the fully N-phase. Strong couplings, (\textbf{g}) and (\textbf{h}): the partial N-branch of $A$ becomes metastable (red symbols) until the continuous SN threshold of layer $A$ merges with its first-order jump (star symbol, Fig.~\ref{fig:3}\textbf{a}). Above this point, the system is purely metastable (\textbf{h}) below the bulk SC-melting threshold, $T_{c,>}$, in its mutual N-phase and mutual SC-phase. 
		}}\label{fig:3}\vspace*{-0.3cm}
\end{figure*}

\paragraph*{\small \underline{Theoretical modeling}} To characterize microscopically the electro-thermal feedback underlying the mutual SN-phase transitions observed in the experiments, we develop a framework (Fig.~\ref{fig:2}\textbf{a}) of thermally-interdependent disordered $2D$-lattices of resistively-shunted Josephson junctions (RSJJs).
In this model (see Methods, M4), the state (SC, intermediate, N) of a given bond, e.g.\ $(i,j)$, is set via a Josephson $I$-$V$ characteristic (Fig.~\ref{fig:2}\textbf{b}) defined by the junction's critical current $I_{ij}^c$ and its normal-state resistance $R_{ij}^n$, whose values depend on the local temperature, $T_{ij}$, reached around it. 
We describe the latter by generalizing the long-celebrated de Gennes relation~\cite{de1976relation} to a local form given by
\begin{equation}
I_{ij}^c(T_{ij})=I_{ij}^c(0)(1-T_{ij}/T_{ij}^c)^2, \label{eq:0}
\end{equation}
\noindent 
where $I_{ij}^c(0)$ is the zero-$T$ critical current of the junction $(i,j)$ and $T_{ij}^c$ is its activation temperature, whose values (caption, Fig.~\ref{fig:2}\textbf{c}) are extrapolated from the experimental data (for details, see Methods, M4). 
To control the degree of disorder in the lattices, we consider a normal distribution with zero mean and standard deviation $\sigma$, in terms of which we generate the junctions' zero-$T$ critical currents, their temperatures $T_c$ and normal-state resistances $R_{ij}^n$ (see Methods, M4). 

When the driving currents $I_{b,A}$ and $I_{b,B}$ injected in the two arrays are kept constant, an increase (decrease) of the cryostat temperature, $T$, controls the layers' SN phase transitions since it reduces (increases) the critical current of single junctions as in Eq.~\eqref{eq:0}. 
This response generally depends on the presence of Joule heating effects which can intertwine the states of the overlapping junctions. 
If thermal couplings are absent, then the local temperatures $T_{ij}^A$ and $T_{ij}^B$ in the arrays coincide with $T$ and the distribution of the critical currents vary {\em homogeneously} with the temperature. 
In this case, local phase perturbations are dampen out and produce rapid transients during which the current optimally redistributes its flow over new isoresistive paths. 
By solving numerically~\cite{ponta2009resistive} the Kirchhoff equations (see Methods, M5) of the isolated RSJJ networks, we find that this redistribution of the currents yields only {\em continuous} SN transitions, with resistive curves (Fig.~\ref{fig:2}\textbf{c}) whose broadness and threshold depend on the degree of disorder of the two arrays (see Fig.~\ref{fig:S1}). 

This scenario drastically changes when the RSJJ networks are thermally interdependent. 
In this case, the states of two overlapping junctions, e.g.\ $a_{ij}$ in layer $A$ and $b_{ij}$ in layer $B$, interact with each other through their local temperatures. 
To include this mutual overheating effect, we consider the junctions' instantaneous dissipation $P_{ij}(t)=R_{ij} I_{ij}^2(t)$, where $R_{ij}$ is the junction's resistive state (Methods, Eq.~\eqref{eq:M1}), so that the local temperatures $T_{ij}^A$ and $T_{ij}^B$ read now as
\begin{equation}\label{eq:1}
\begin{aligned}
\begin{pmatrix}
T_{ij}^A(t) \\ T_{ij}^B(t)
\end{pmatrix} 
=T+ &\,
\gamma
\begin{pmatrix}
0 & 1 \\
1 & 0
\end{pmatrix}
\begin{pmatrix}
R_{ij}^A I_{ij, A}^2(t-1) \\ 
R_{ij}^B I_{ij, B}^2(t-1)
\end{pmatrix} +\\
+&\,
\gamma'
\begin{pmatrix}
1 & 0 \\
0 & 1
\end{pmatrix}
\begin{pmatrix}
\sum_{k\in\partial(i,j)} R_{k}^A I_{k, A}^2(t-1) \\ 
\sum_{k\in\partial(i,j)} R_{k}^B I_{k, B}^2(t-1) 
\end{pmatrix},
\end{aligned}
\end{equation}
\noindent 
where $\gamma$ and $\gamma'\,\mathrm{[KW^{-1}]}$ are, respectively, the thermal conductances of the coupling medium and of the networks' substrate. 
Eq.~\eqref{eq:1} is general and it includes the overheating of $(i,j)$ due to the dissipation produced by its $\partial(i,j)$ neighbouring junctions. 
In our samples, however, the thickness of the $\mathrm{Al_2O_3}$ layer is roughly $2$ orders of magnitude (Fig.~\ref{fig:1}\textbf{b}, inset) smaller then the lattice spacing within each array and its thermal conductivity~\cite{berman1951thermal} is about 50 times larger then the $\mathrm{SiO}_2$~\cite{moore2014emerging} substrate (respectively, $10\mathrm{W/mK}$ vs.\ $0.2\mathrm{W/mK}$), suggesting that $\gamma'\ll \gamma$, i.e.\ junctions are weakly thermally-interdependent within the layers. 
Under this condition, the iterative two-interactions interplay set between the layers by Eq.~\eqref{eq:0} and Eq.~\eqref{eq:1} yields an {\em adaptive} and {\em heterogeneous} response of the critical currents to local thermal fluctuations that describes mathematically the electro-thermal runaway effect triggered by cross-layers interdependent couplings. 

To study the system in this configuration, we solve numerically the thermally-coupled Kirchhoff equations (Methods, Eqs.~\eqref{eq:M1}--\eqref{eq:M3}) set by the process described above for disordered lattices whose physical properties (caption, Fig.~\ref{fig:2}\textbf{c},\textbf{d}) match those in the experiments. 
During each stage in the cascade of overheatings, we compute the junctions' current, their electronic state and the power dissipated in order to track their spatio-temporal evolution (see Fig.~\ref{fig:S2},\,\ref{fig:S3} and Supplementary Movies S1-S2). 
For driving currents $I_{b,A},\,I_{b,B}\geq 15\mu\mathrm{A}$, the system enters a regime of mutual first-order SN transitions (Fig.~\ref{fig:2}\textbf{d}) accompanied by different microscopic dynamics. 
Fig.~\ref{fig:2}\textbf{e},\textbf{f} show, in particular, the bulk resistances of the layers and the local power-dissipated when relaxing the system from the deep SC-phase to the full N-phase at a temperature $T$ slightly above the first-order SN threshold, $T_{c,>}$. 
As seen in Figs.~\ref{fig:2}\textbf{e},\textbf{f}, above $T_{c,>}$ the mutual SC-phase undergoes a long-lived (``plateau'') relaxation characterized by cascades of N-switching junctions, whose duration $\tau\sim(T-T_{c,>})^{-\zeta}$ with $\zeta\simeq0.65$ diverges at the critical threshold (Fig.~\ref{fig:S4}\textbf{c},\textbf{d}) as in other critical bulk-melting processes above first-order transitions~\cite{binder1987theory} \!\!. 
In the cooling direction, the evolution from the mutual N-phase to the mutual SC-phase exhibits an analogous metastable stage (Fig.~\ref{fig:S4}\textbf{a},\textbf{b}) whose duration diverges at the bulk N-to-SC threshold, $T_{c,<}$, as $\tau\sim(T_{c,<}-T)^{-\zeta}$, now with exponent $\zeta\simeq0.5$. 
The different critical exponents (Fig.~\ref{fig:2}\textbf{g}) of the metastable lifetime at the first-order thresholds can be taken as proxies of the underlying cascading processes, indicating that SC-nuclei grow faster then N-nuclei. 
During the heating plateau, this can be explained in terms of the pinning of the interfaces between SC-clusters and N-nuclei which halts their growth, while the smaller (in fact, mean-field~\cite{krzakala2011melting1}) exponent at the cooling plateau hints at the sudden merging of thermally-suppressed SC-clusters. 

\paragraph*{\small \underline{Mean-field solution}} 
To further corroborate our findings, we develop an analytical mean-field (MF) solution of the thermally-interdependent Kirchhoff equations under the two-interactions interplay set by Eq.~\eqref{eq:0} and Eq.~\eqref{eq:1}. 
We build our MF-solution on the long-celebrated Halperin-Nelson (HN)\vspace*{-0.05cm} formula~\cite{halperin1978theory} $R_{HN}(T)=R_0\mathrm{exp}\{-\beta\big(T-T_{c}(I_b)\big)^{-1/2}\}$---where $R_0$, $\beta$ and $T_c(I_b)=T_{c,0}-\omega I_b$ are material parameters---which characterizes the resistance of a $2D$ superconductor slightly above its continuous SN transition. 
We advance a global-coupling hypothesis (see Methods, M6, for details on its validity) by adopting an all-to-all network of thermal dependency couplings between the layers so that the RSJJ arrays interact through their collective phases. 
This is done by replacing the local quantities in Eq.~\eqref{eq:1} by their global counterparts, which coarse-grains the system of $4L(L-1)$ local temperatures into $2$ global ones. 
Since the length of dependency links is random (i.e.\ $\sim\mathcal{O}(L)$), the global overheating at the $t$-th stage of the cascade on layer $\mu$ due to the power dissipated by layer $\mu'$ can be computed via the HN-resistance of $\mu'$ at the effective temperature induced by $\mu$ on $\mu'$ at the previous stage, $t-1$, and so forth in a recursive fashion. 
We can then represent the evolution of overheating cascades via the recursion sequence of adaptive global temperatures: 
\begin{equation}\label{eq:3}
T_{\text{eff},t}^{\mu\leftarrow\mu'}= T + \gamma R_{HN}^{\mu'}(T_{\text{eff},t-1}^{\mu'\leftarrow\mu}\big)I_{b,\mu'}^2,\quad n=1,2,\dots
\end{equation}
\noindent
for $\mu'\neq \mu$ and $\mu,\mu' =A,\,B$, with the initial seed $T_{\text{eff},0}^{A\leftarrow B}\equiv T$. 
In the limit $t\to\infty$, the fixed points of Eq.~\eqref{eq:3} yield a system of self-consistent equations for the mutual bulk resistances
\begin{equation}\label{eq:4}
\begin{cases}
R_{HN}^{A}(T)=R_0^A e^{-\beta_A/\sqrt{T-T_{c}^A(I_{b}^A)-\gamma R_{HN}^{B}(T)}},\\
R_{HN}^{B}(T)=R_0^B e^{-\beta_B/\sqrt{T-T_{c}^B(I_{b}^B)-\gamma R_{HN}^{A}(T)}},
\end{cases}
\end{equation}
\noindent 
which can be solved numerically for suitable choices of the material-dependent parameters (caption, Fig.~\ref{fig:3}). \\
\indent 
We find that the MF-theory nicely agrees with the numerical results (Fig.~\ref{fig:3}\textbf{a}) and correctly captures the phenomenology of mutual SN-phase transitions observed in the experiments within the accessible range of parameters (see also Figs.~\ref{fig:S5}, \ref{fig:S7}). 
Depending on the values of the driving currents, we identify 3 main coupling regimes of increasing strengths (blue-to-red color bar, Fig.~\ref{fig:3}\textbf{a}): $i)$ {\em weak interdependence} ($I_b\lesssim17$), where both layers undergo continuous SN-transitions (Fig.~\ref{fig:3}\textbf{c}, \textbf{d}); $ii)$ {\em moderate interdependence} ($17\lesssim I_b\lesssim40$), where two-steps (continuous and first-order) transitions are observed (Fig.~\ref{fig:3}\textbf{e}--\textbf{g}); $iii)$ {\em strong interdependence} ($I_b\gtrsim40$) where the system undergoes only mutual first-order SN transitions (Fig.~\ref{fig:3}\textbf{h}).
In the intermediate regime, in particular, the continuous SN-transition of layer $A$, i.e.\ the one having the lowest bulk critical temperature, is reversible only for $I_b\lesssim20$ (orange and red symbols, Fig.~\ref{fig:3}\textbf{e},\,\textbf{f}) and it is always followed by a mutual first-order jump to the full N-phase. 
For $I_b\gtrsim20$, layer $A$ enters a marginally stable partial N-phase (red symbols, Fig.~\ref{fig:3}\textbf{g} and orange full line in Fig.~\ref{fig:3}\textbf{a}) whose threshold rapidly converges to the bulk SN-heating one (dashed red curve, Fig.~\ref{fig:3}\textbf{a}) when $I_b$ is increased. 
Instead, when cooling the system from its mutual full N-phase, for $I_b\gtrsim20$ both layers undergo coupled first-order NS (i.e.\ N-to-SC) phase transitions whose thresholds (blue dashed curve, Fig.~\ref{fig:3}\textbf{a}) rapidly decrease for increasing currents. 
In particular, when $I_b\gtrsim40$ (star symbol, Fig.~\ref{fig:3}\textbf{a}) the partial-N branch vanishes and the two arrays becomes fully thermally-interdependent (see also Fig.~\ref{fig:S5}). 
In this regime, the MF-theory predicts a zero-temperature mutual metal ground state that coexists with the mutual SC-one (phase $\mathrm{VII}$, Fig.~\ref{fig:3}\textbf{a}) in a thermally-bistable electronic state. 
A full classification of the mutual phases in the system (Fig.~\ref{fig:3}\textbf{a}, inset) is given in the caption to Fig.~\ref{fig:3}\textbf{a}. 

\paragraph*{\small \underline{Summary}} 
At its heart, physics is about identifying natural phenomena, developing models harnessing the laws that govern them, and then adopting those models to control and/or to predict novel behaviors. Laboratory-controlled experiments are crucial in this regard, as they nourish that back-and-forth mechanism between abstract ideas and realistic constraints that drives the discovery of new perspectives and frontiers. Over the last decade, the lack of experimental realizations of interdependent systems has disabled this virtuous interplay, constraining our understanding of their complexity within the realm of mathematical modeling. 
The system of thermally-coupled superconductors developed and studied here roots the theory of interdependent networks into the physical laboratory, opening a Pandora box of scientific challenges. 
For example, instead of being thermal, dependency links may emerge as magnetic, capacitive or inductive feedbacks in other physical systems---e.g.\ coupled BKT-vortices between two layers of $2D$ magnets~\cite{huang2017layer, gibertini2019magnetic} \!\!---whose realization would foster the development of further interdependent materials embodying the ``two-interactions'' paradigm. 
Besides raising challenges in the development of predictive theories of interdependent materials, 
the striking discontinuity of the SN transition in coupled networks demonstrated in this work offers the unprecedented opportunity of designing innovative technologies, like ultra-sensitive sensors~\cite{shurakov2015superconducting} and multi-stack memory devises~\cite{meijer2008wins} \!\!, that exploit the spontaneous emergence of mutual macroscopic phases due to the back-and-forth cascade of microscopic perturbations. }


\FloatBarrier
\bibliographystyle{unsrt}
\bibliography{RSJJ.bib}

\renewcommand\thesection{M\arabic{section}}
\renewcommand\thesubsection{M\arabic{section}.\arabic{subsection}}
\setcounter{section}{0}
\setcounter{equation}{0}
\setcounter{figure}{0}
\renewcommand{\theequation}{M\arabic{equation}}
\renewcommand{\thefigure}{M\arabic{figure}}

{\footnotesize
\section*{Methods}

M1) \emph{\underline{Sample preparation}}. 
The interdependent superconducting system (see the schematic representation in Fig.~\ref{fig:1}\textbf{b}, main text) was prepared as follows. \textbf{1}) On a  ($\mathrm{Si}/\mathrm{Si}\mathrm{O}_2$) substrate we e-beam evaporated a thin film of $50\mathrm{nm}$ a:InO with partial oxygen pressure ($6$--$8\mu\mathrm{Torr}$), resulting in disordered superconductor with a bulk critical temperature $T_c\simeq 3\mathrm{K}$. The layer was patterned to form a network consisting of $31X31$ stripes, each one being $4\mu\mathrm{m}$ wide and $720\mu\mathrm{m}$ long (Fig.~\ref{fig:1}\textbf{b}, inset), thus resulting in a superconducting lattice composed of segments with dimensions $4\times20\mu\mathrm{m}$ and height $\sim 50\mathrm{nm}$.    \textbf{2}) For the electrically insulating medium, we evaporated a thin film of $100$--$150\mathrm{nm}$ of $\mathrm{Al}_2\mathrm{O}_3$ on top of the network  at high partial $O_2$ pressure in order to achieve a pinhole free film. 
\textbf{3}) On the top of the $\mathrm{Al}_2\mathrm{O}_3$ layer, we evaporated a second superconducting network sample, perfectly overlapping the first one (Fig.~\ref{fig:1}\textbf{b}, inset). 
\textbf{4}) We then fabricated two Au contacts of 4nm thick $\mathrm{Cr}$ + $35\mathrm{nm}\mathrm{Au}$ at the edges of each network in order to enable independent transport measurements. 
The adoption of $\mathrm{Al}_2\mathrm{O}_3$ as a coupling medium is motivated by its strong electrical insulating properties and relatively large thermal conductivity ($\lambda_{\mathrm{Al}_2\mathrm{O}_3} \sim 10\mathrm{W}/mK$ at $T\simeq 3K$, from Ref.~\cite{berman1951thermal}), which enables to realize a cross-layers heat-transfer without the hopping electrons. 

M2) \emph{\underline{Thermal conductances}}. 
We estimate the inter-layer thermal conductance, $\gamma'[\mathrm{KW}^{-1}]$ in Eq.~\eqref{eq:1}, by considering the resistive curve at the minimal current---here, $I_{b,min}=1\mu\mathrm{A}$ (Fig.~\ref{fig:S0}\textbf{a}, inset)---as a reference. We then measure the value of $R(T)$ for the resistive curves at larger $I_b$ values at a certain cryostat temperature $T_{cryo}$ ($T_{cryo}=3\mathrm{K}$ in Fig.~\ref{fig:S0}\textbf{a}) and determine the appropriate temperature, $T_{ref}(I_b)$, for these resistances in the reference curve. We calculate $\Delta T = T_{cryo}-T_{ref}$ for each curve and plot it against the power, $P$, applied to the network: the slope $\mathrm{d}\Delta T/\mathrm{d}P$ yields an estimate of the thermal conductance $\gamma'$ within each layer, as shown in Fig.~\ref{fig:S0}\textbf{b} for an a:InO network on a $\mathrm{SiO}_2$ substrate. We find that the thermal conductance within the superconducting network is then $\gamma'\simeq 10^5\mathrm{KW}^{-1}$. Since the distance between superposed junctions is, roughly, $100$ times smaller than the distance between the centres of orthogonal junctions (Fig.~\ref{fig:1}\textbf{b}, inset), we conclude that the cross-layers thermal conductivity is $\gamma\simeq 100 \gamma'$. 

M3) \emph{\underline{Measurements}}.  
We performed $\mathrm{DC}$-transport measurement using a Keithley $2410$ sourcemeter and a Keithley $2000$ multimeter for each network. The cryostat temperature was tuned via a LakeShore $330$ using a $25\Omega$ heater and a $DT$-$670$ thermometer was placed inside the cryostat. We started by measuring the global {\em sheet} resistance of each superconducting array with adiabatic heating-cooling cycles in the temperature range base to $10\mathrm{K}$ for different values of the driving current, $I_b$. After characterizing the phase diagrams of each isolated array (Fig.~\ref{fig:3}\textbf{b}, inset), we checked the absence of shorts between the layers by measuring the junction resistance between each pair of cross contacts. The cross-layers couplings are created by passing the same current within both layers simultaneously, thus generating dependency links sustained by heat-transfer. $\mathrm{DC}$-transport measurements were then performed in the thermally-interdependent setup with adiabatic heating-cooling cycles for the same currents as in the isolated case, yielding the curves in Fig.~\ref{fig:1}\textbf{f} and the coupled phase diagram in Fig.~\ref{fig:3}\textbf{b}. 

M4) \emph{\underline{RSJJ-Model of disordered superconductors}}. To characterize the SN transitions observed in the experiments, we model each disordered superconductor via a disordered $2D$-lattice of resistively-shunted Josephson junctions (RSJJs). Isolated networks of RSJJs undergo {\em continuous} SN-phase transitions at low temperatures that, in the limit of large tunneling conductances (i.e.\ $g\gg1$), are generally independent on the ratio between the Josephson $E_J$ and the Coulomb $E_C$ energies~\cite{orr1986global, chakravarty1986onset, chakravarty1988quantum} \!\!. In this regime, each junction's state can be characterized by the value of its normal-state resistance, $R_n(T)$, and by its critical current, $I_c(T)$, which generally depend on the ratio between the temperature $T$ of the cryostat and the junction's SN-activation threshold $T_{c}$. When dealing with ordered superconducting arrays, the letter quantities satisfy in the so-called dirty limit~\cite{abraham1982resistive, lobb1983theoretical} the Ambegaoakar-Baratoff relation~\cite{josephson1962possible, ambegaokar1963tunneling} $I_c(T)R_n=\frac{\pi }{2e}\Delta(T)\mathrm{tanh}(\Delta(T)/2k_BT)$, where the energy gap follows the BCS mean-field spectral relation $2\Delta(T)\simeq\alpha k_B T_c$ with $\alpha\simeq3.53$. In disordered superconductors, on the other hand, disorder-induced spatial inhomogeneities of the SC-state break the ideal BCS scheme in the above, yielding striking phenomena~\cite{dubi2007nature} like non-monotonic variations of the sheet resistance~\cite{baturina2007localized} \!\!, suppression of $T_c$ towards zero~\cite{sacepe2008disorder} and large values~\cite{sacepe2011localization} of the spectral gap ratio $\Delta(T)/T_c$. When modeling these networks via RSJJs, an Arrhenius activation law at low temperatures~\cite{ponta2013superconducting} is invoked to include the presence of large resistive areas due to the emergence of insulating islands. The a:InO samples fabricated in the present work, however, have bulk SN-thresholds large enough to ensure that junctions rarely undergo a metal-insulator (MI) transition. In light of this, we consider a model of RSJJ with only three electronic states: superconducting (SC), intermediate (IM) and normal-metal (N), defined according to the Josephson $I$-$V$ characteristic displayed in Fig.~\ref{fig:2}\textbf{b}. Hence, the junction's resistance is defined piecewise as: 
\begin{equation}
        R_{ij} = 
        \begin{cases}
            R_{\epsilon}, & \text{if } V_{ij} < R_{\epsilon} I_{ij}^c (T) \text{ (SC)},\\
            R_{ij}^n, & \text{if } V_{ij} > R_{ij}^n I_{ij}^c (T) \text{ (N)},\\
            V_{ij} /  I_{ij}^c (T), & \text{otherwise} \text{ (IM)},
        \end{cases}
        \label{eq:M1}
\end{equation}
\noindent 
where $R_{\epsilon}$ is the resistance in the SC-state ($R_\epsilon=10^{-5} \Omega$ in the simulations, see M5 for details) and $V_{ij}$ is the potential drop measured at the junction's ends. For the critical currents, we propose a local generalization of the de-Gennes relation, i.e.\ Eq.~\eqref{eq:0}, where $I_{ij}^c(0)$ is the junction's critical current at $T=0$. We control the degree of disorder in the arrays by considering a quenched normal distribution $\mathpzc{X}_{ij}\in\mathcal{N}(0,\sigma)$---where variables match the junctions' labels in each array---with zero mean and variance $\sigma$ as a generator for the other system's observables. In particular, we define $I_{ij}^c(0)=I_0^c(1+\mathpzc{X}_{ij})$, $T_{ij}^c=T_c(1+\mathpzc{X}_{ij})$ and $R_{ij}^n=\rho R_q(1+\mathpzc{X}_{ij})$,\vspace*{-0.05cm} where $R_q\equiv h/4e^2\simeq6.45\mathrm{k}\Omega$ is the quantum resistance for pairs, so that junctions with a large zero-$T$ threshold have a comparably large critical temperature and normal resistance. The values of $I_0^c$, $T_c$ and $\rho$ can be extrapolated from the experimental data; $T_c$, in particular, can be found by fitting to the resistive curves the Aslamazov–Larkin (AL) correction~\cite{aslamazov1996effect, baturina2004superconductivity} or, similarly, the Halpering-Nelson (HN) relation~\cite{halperin1978theory, halperin1979resistive} adopted in the main text. Best fitted values for $I_0^c$, $T_c$ and $\rho$ are listed in the caption to Fig.~\ref{fig:2}. 

M5) \emph{\underline{Thermally-coupled Kirchhoff equations}}. 
To characterize the mutual SN-phase transitions reported in the experiments, we have developed a model of thermally-coupled RSJJs networks with local thermal couplings sustained by the heat dissipation of single junctions. Alike simulations in interdependent networks~\cite{gao2012networks} \!\!, numerical solutions for the mutual order parameter (here, the global sheet resistance, $R$) can be obtained recursively by making the layers to adaptively interact through their isolated behaviors~\cite{ponta2009resistive} \!\!. In our model of thermally-interdependent RSJJs networks, this is achieved by solving the Kirchhoff equations of each array under the adaptive effect set by the ``two-interactions'' interplay between Eq.~\eqref{eq:0} and Eq.~\eqref{eq:1}. We consider therefore two layers, $A$ and $B$, each being a $2D$ lattices with linear size $L$, whose left and right boundaries are connected to an external super-node (source) where the bias current is injected and to the ground, respectively. Each junction has a Josephson $I$-$V$ characteristic with $R_{ij}$ defined as in Eq.~\eqref{eq:M1}, where we assume $R_{\epsilon}=10^{-5}\Omega$ for both the arrays and mean normal-resistance $R_n=\rho R_q$ with $\rho_A=1.24$ and $\rho_B=0.77$. We initiate the algorithm by randomly assigning two vector potentials $W_\mu$ with $\mu=A,B$ with same values for all junctions at the $0$-th iteration. When starting from the mutual SC-state, the junctions' resistances in both layers are set as\vspace*{-0.05cm} $R_{ij}^A = R_{ij}^B = R_{\epsilon}$, whilst $R_{ij}^A=R_{ij,A}^n$ and $R_{ij}^B=R_{ij,B}^n$ when the layers start from their mutual N-phase. The algorithm evolves iteratively as follows: 
\begin{itemize}
\item[$1)$] at the $t$-th stage ($t\geq1$) of the overheating cascade, the local effective temperatures, Eq.~\eqref{eq:1}, are computed using the resistances and the local currents found at the stage $(t-1)$; 
\item[$2)$] the critical currents $I_{ij}^c(T)$ are updated via Eq.~\eqref{eq:0}, and their resistive state is determined via Eq.~\eqref{eq:M1} after computing the potential drop $V_{ij,t}$ from the vector $W_t$; 
\item[$3)$] the (symmetric) conductance matrices $\bar{G}_\mu$ with $\mu=A,B$ are generated via the junctions' resistances in $2)$ with entries 
\begin{equation}
        G_{ij} = 
        \begin{cases}
            0, & \text{if}\quad (i,j)\notin E \\
            -1 / R_{ij}, &  \text{if}\quad (i,j)\in E \\
            \sum_{k \in \partial i} 1 / R_{ik}, &  \text{if}\quad i = j
        \end{cases}
        \label{eq:M2}
\end{equation}
where $E$ is the set of edges in each arrays and $\partial i$ the set of nearest neighbours of node $i$; 
\item[$4)$] the potential vectors, $W_{\mu, t+1}$, are updated by solving numerically the Kirchhoff matrix equations 
\begin{equation}
    \begin{cases}
    \bar{G}_t^A \cdot W_{t+1}^A = I_{inj}^A \\
    \bar{G}_t^B \cdot W_{t+1}^B = I_{inj}^B
     \end{cases}\label{eq:M3}
\end{equation}
\noindent 
where $(\,\cdot\,)$ is the matrix product and $I_{inj}^\mu$ is the vector of total currents injected to each node at every stage, whose elements are always zeroes except for the first entry (the super-node) which equals the driving current $I_b^\mu$ with $\mu=A,B$; 
\item[$5)$] the global sheet resistances of each array are then calculated as $R_{t+1}^\mu=W_{t+1}^\mu/I_b^\mu$ with $\mu=A,B$. 
\end{itemize}
\noindent 
The steps 1)--5) are recursively repeated yielding a sequence of pairs of vector potentials:
$\{(W_0^\mu, W_0^B), \dots, (W_t^A, W_t^B),\dots\}$, whose convergence is verified as soon as the mutual error
$$
\delta W = \sum_{\mu=A,B}\bigg|1-\frac{W_t^\mu}{W_{t+1}^\mu}\bigg| 
$$
\noindent 
becomes smaller than a numerical precision $\varepsilon_{min}$. In the simulations carried on in the present work, we used $\varepsilon_{min} = 10^{-5}$; we verified that higher precision thresholds do not alter the phase diagram of decoupled and thermally-interdependent networks. 

M6) \emph{\underline{Validity of the mean-field hypothesis}}. 
Non-locality\vspace*{-0.05cm} is an essential feature for the large-scale propagation of cascades~\cite{motter2017unfolding} \!\!. In our model of thermally interdependent $2D$ SC-networks, the random redistribution of the currents after the state-switch of single junctions propagates non-locally local phase perturbations, setting an {\em effective} long-range feedback within each layer. Recent findings on percolation in interdependent spatial networks~\cite{li2012cascading, danziger2016effect, gross2021interdependent} unveil that randomly interdependent lattices (i.e.\ coupled $2D$ grids with long-range dependency links) and multiplex disordered lattices (i.e.\ coupled spatially embedded networks with long-range connectivity links and dependency links between overlapping nodes) are physically equivalent, featuring the same equilibrium phases and dynamical regimes. This equivalence finds solid grounds in the mapping~\cite{bonamassa2021realizing} we have recently discovered between percolation in $K>2$ randomly interdependent networks and the onset of hard-fields in the one-step-replica-symmetry-breaking solution of the random $(K+1)$-\textsc{xorsat} problem, i.e.\ with the ground state of ferromagnetic $(K+1)$-spin models on random hypergraphs. Since an interdependent link between nodes interacting with their nearest neighbours via pairwise couplings maps exactly onto a hyper-edge made by triads of the two dependent nodes and their nearest neighbours, a source of long-rangedness either on the dependency links or on the connectivity links yields statistically equivalent structures (i.e.\ hypergraphs with triads between 2 nodes at short-range distance and 1 randomly chosen node within the arrays). 
In light of this, the mean-field hypothesis advanced in the main text can be read as the completely random version of the above, which does not alter the main phenomenology of first-order transitions and cascade of failures observed in fully random interdependent networks~\cite{gao2012networks} \!\!.

\newpage
\textbf{\small Code availability}. Source codes and videos showing the states of resistors, their currents and the power dissipated in both layers during the transition can be freely accessed at the GitHub repository: \url{https://github.com/BnayaGross/Interdependent-SC-networks}.

\textbf{\small Data availability}.
All data supporting our findings are available from the corresponding author upon reasonable request.

\textbf{\small Acknowledgments}. 
S.H.\ acknowledges financial support from the ISF, the China-Israel SF, the ONR, the BIU Center for Research in Applied Cryptography and Cyber Security, the EU project RISE, the NSF-BSF Grant No. 2019740, and the DTRA Grant No. HDTRA-1-19-1-0016. I.B., A.F.\ and S.H.\ acknowledge partial support from the ITA/ISR grant ``\textsc{Explics}''. 

\textbf{\small Author contributions}. 
I.B., A.F.\ and S.H.\ initiated and designed the research. 
M.L., I.V.\ and A.F.\ fabricated the samples, carried on the experiments and collected the data. 
I.B.\ and B.G.\ developed the modeling and the adaptive algorithm for solving the thermally-interdependent Kirchhoff equations. 
B.G.~designed the codes and carried out the numerical simulations with contributions from I.B.. 
I.B.\ developed the mean-field theory. 
I.B.~was the leading writer of the manuscript with contributions from B.G., S.H.\ and A.F.. 
A.F.\ and S.H.~supervised the research. 
All authors critically reviewed and approved the manuscript. }

\newpage
\onecolumngrid

\renewcommand\thesection{S\arabic{section}}
\renewcommand\thesubsection{S\arabic{section}.\arabic{subsection}}
\setcounter{section}{0}
\setcounter{equation}{0}
\setcounter{figure}{0}
\setcounter{page}{1}
\renewcommand{\theequation}{S\arabic{equation}}
\renewcommand{\figurename}{\textsc{Extended Data Figure}}
\renewcommand{\thefigure}{ED\arabic{figure}}
\renewcommand{\thepage}{S\arabic{page}}

\newpage

\begin{quote}
\centering 
{\large \bf Interdependent superconducting networks}\vspace*{+0.1cm}\\
{\normalsize (\underline{\textsc{Extended Data}})}\vspace*{+0.25cm}\\
{\textsc{I.\ Bonamassa$^{1,2}$, B.\ Gross$^{1}$, M.\ Laav$^{1}$, I.\ Volotsenko$^{1}$, A.\ Frydman$^{1}$, S.\ Havlin$^{1}$}}\\
{\small \em $^{1}$Department of Physics, Bar-Ilan University, 52900, Ramat-Gan, Israel}\\
{\small \em $^{2}$Department of Network and Data Science, CEU, Quellenstrasse 51, A-1100 Vienna, Austria}\\
{\small (Dated: \today)}
\end{quote}

\begin{figure}[h!]\vspace*{+5cm}
\centering
	\sbox\mysavebox{
	\includegraphics[width=0.85\linewidth]{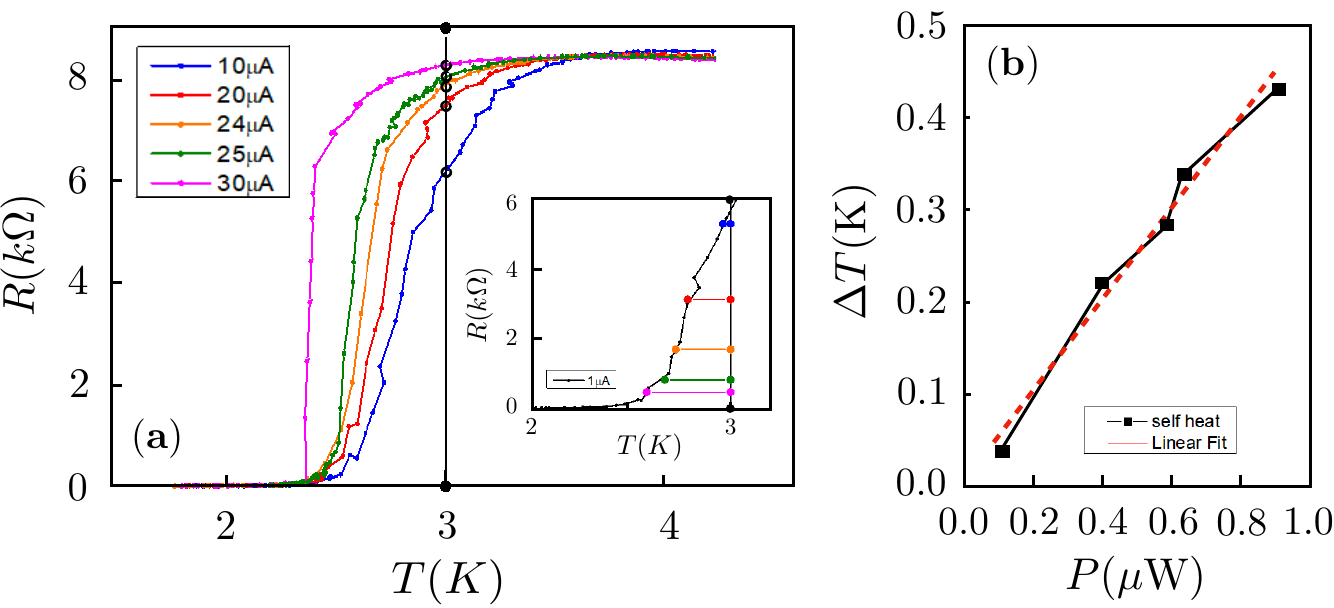}}%
	\usebox\mysavebox
  	\par
  	\begin{minipage}{\wd\mysavebox}\vspace*{+0.3cm}
		\caption{\small \textbf{Estimating the intra-leyer thermal conductance}. (Color online) 
		(\textbf{a})~Collection of the resistances at $T=3\mathrm{K}$ for different resistive curves at increasing driving currents, $I_b$. (Inset) Temperatures associated to the resistances collected in (\textbf{a}) on the reference curve at $I_b=1\mu\mathrm{A}$. 
		(\textbf{b})~Temperature difference $\Delta T$ as a function of the power applied to the network whose best fit yields the slope $\gamma'\sim10^6\mathrm{KW}^{-1}$. } \label{fig:S0}
	\end{minipage}
\end{figure}

\newpage
\begin{figure}[h!]\vspace*{+4cm}
	\centering
	\sbox\mysavebox{
	\includegraphics[width=0.85\linewidth]{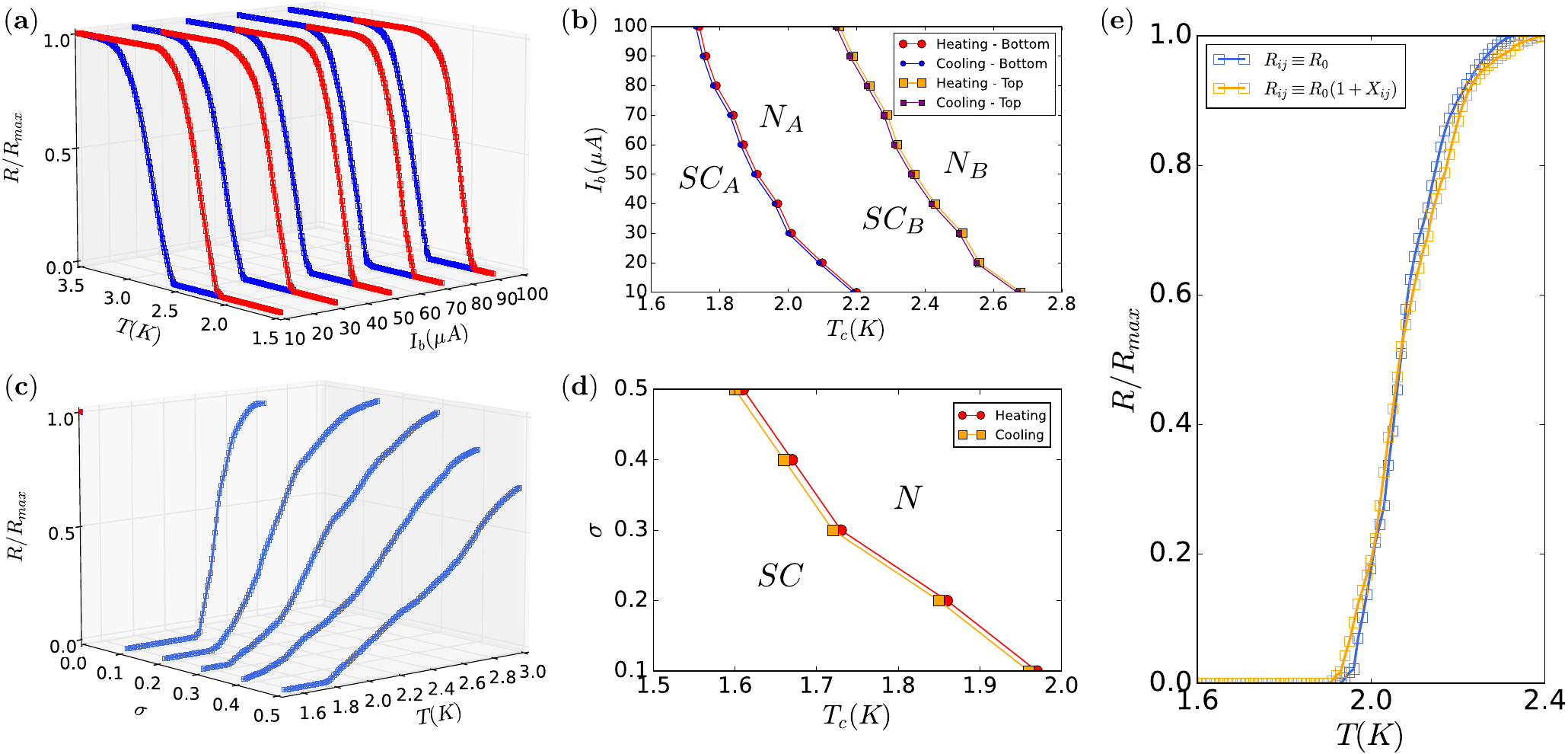}}%
	\usebox\mysavebox
  	\par
  	\begin{minipage}{\wd\mysavebox}\vspace*{+0.3cm}
		\caption{\small (Color online) {\bf Numerical resistive transitions in RSJJ networks.} 
		(\textbf{a}) Continuous SN-transitions in the normalized bulk resistance of two {\em independent} RSJJ networks (blue and red symbols). Numerical data are obtained by solving the Kirchhoff equations Eq.~\eqref{eq:M2}, \eqref{eq:M3} (see Methods, M4 and M5) in lattices with $N=2L(L+1)$ junctions and $L=30$. Red (blue) symbols correspond to networks of JJs with material parameters $I_{c,0}^A=48\mu\mathrm{A}$, $T_c^A=2.4\mathrm{K}$ and $\rho_A=1.24$ ($I_{c,0}^B=58\mu\mathrm{A}$, $T_c^B=2.9\mathrm{K}$ and $\rho_B=0.77$) and normal distribution of disorder with variance $\sigma=0.1$ (see Methods, M5 for details on the parameters' definition). The results show that the SN transition, in absence of thermal couplings within or between layers is always continuous for increasing values of the driving current $I_b$. 
		(\textbf{b}) Phase diagram characterizing the cooling and heating thresholds where the resistive curves in (\textbf{a}) undergo a continuous SN phase transition. Notice the slight decrease in the bulk $T_c$'s due to the dependence of the junctions' critical currents on the critical temperatures (see Eq.~\eqref{eq:1} in the main text). 
		(\textbf{c}) Continuous SN-transitions in the normalized bulk resistance of a single RSJJ array with $L=20$, $I_{c,0}=48\mu\mathrm{A}$, $T_c=2.4\mathrm{K}$ and $\rho=1.24$. Different curves describe the behavior reported at increasing levels of disorder, which can be suitably controlled in the model by means of the variance $\sigma$ characterizing the normally distributed variables $\mathcal{X}_{ij}\in\mathcal{N}(0,\sigma)$ (Methods, M4). Larger values of $\sigma$, corresponding to larger levels of disorder, result in broader continuous SN-transitions. 
		(\textbf{d}) Phase diagram displaying the cooling and heating critical thresholds characterizing the continuous SN-transitions in (\textbf{c}). 
		(\textbf{e}) Effect of distributed normal state resistances (yellow symbols) vs.\ identical normal-state resistances (blue symbols) in a single RSJJ network. Numerical results are obtained for a network with $L=30$, $\sigma=0.1$, $T_c=2.4$ and $I_{c.0}=48\mu\mathrm{A}$. Both resistive curves and bulk critical thresholds are nearly identical. }\label{fig:S1}
	\end{minipage}
\end{figure}

\newpage
\begin{figure}[h!]\vspace*{+2.5cm}
	\centering
	\sbox\mysavebox{
	\includegraphics[width=0.85\linewidth]{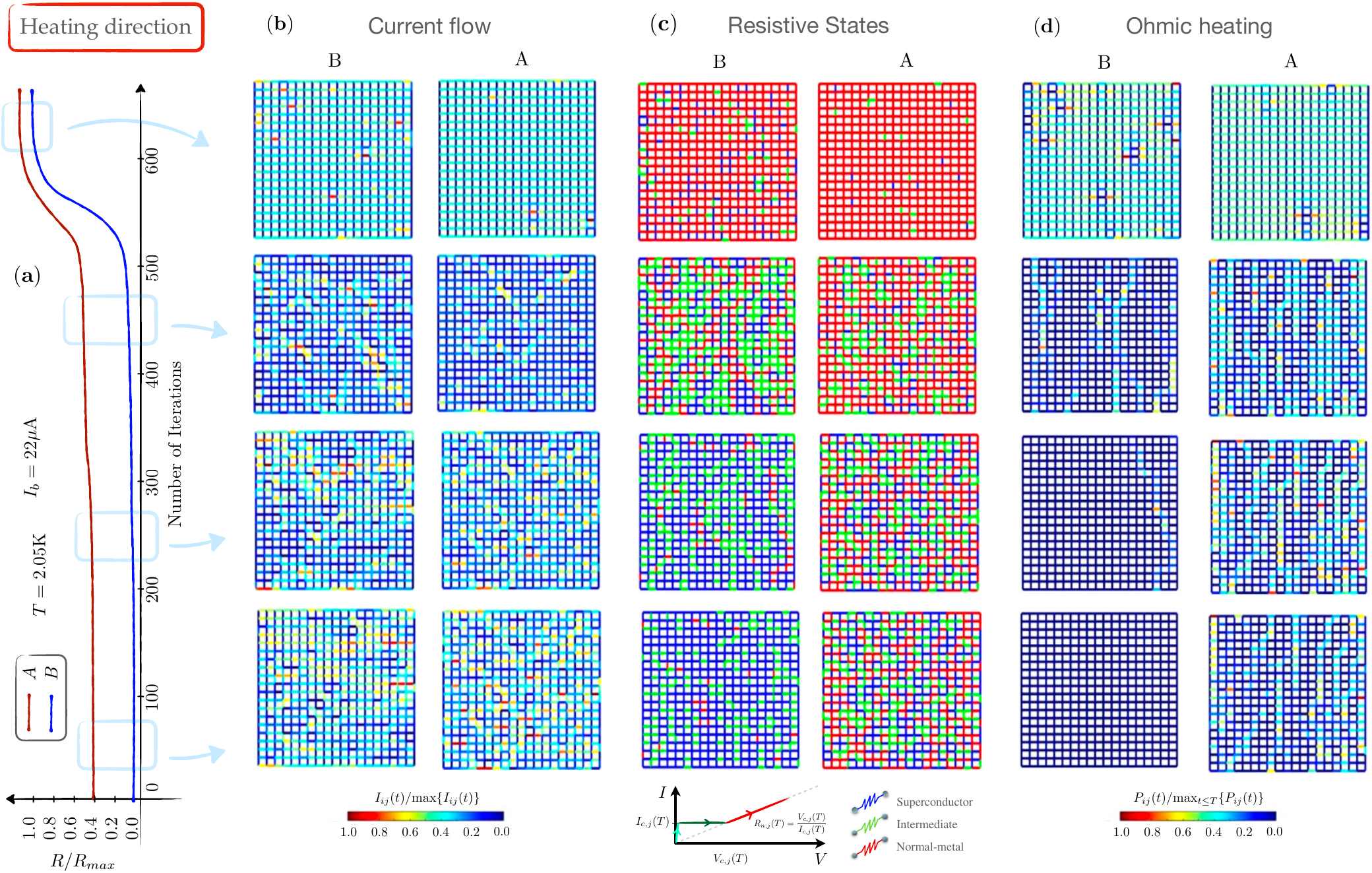}}%
	\usebox\mysavebox
  	\par
  	\begin{minipage}{\wd\mysavebox}\vspace*{+0.3cm}
		\caption{\small (Color online) {\bf Heating plateau and junctions' state evolution.} 
		(\textbf{a}) Evolution of the metastable mutual SC-phase reported in the bulk resistance of $2$ thermally-interdependent RSJJ networks deep quenched slightly {\em above} the bulk SC-to-N first-order transition threshold, $T_{c,>}=2.08\mathrm{K}$. Notice that the more disordered layer (red curve) reaches the partial N-state exponentially rapidly, while the less disordered layer (blue curve) remains in a marginally stable SC-state for longer time. During the evolution, junctions in $A$ switching to the N-state become local dissipating hot-spots, warming up the overlapping junctions in layer $B$, thus cascading the SC-to-N switches of junctions between the layers. This is nicely reflected in the evolution of the junctions' state (see also Fig.~\ref{fig:S4}\textbf{c},\textbf{d}), whose snapshots along the evolution are shown in the columns (\textbf{b})--(\textbf{d}). 
		(\textbf{b}) Stroboscopic snapshot of the instantaneous junctions' current evolution (increasing time from bottom to top) from the mutual SC-phase to the mutual N-phase. Notice the patterns of current redistribution emerging when approaching the mutual N-phase (see also Supplementary Movies S1 and S2), whose patterns resemble those reported in the stress-energy release at the jamming transition of hard-discs~\cite{banigan2013chaotic} \!\!. 
		(\textbf{c}) Evolution of the junctions' resistive states (red = N, green = intermediate, blue = SC; see Methods, M4) during the heating plateau: notice that N-islands progressively percolate the system while SC-paths become rarer and rarer along the electro-thermal runaway due to the thermal-feedbacks and the current redistribution (interplay between Eq.~\eqref{eq:0} and Eq.~\eqref{eq:1} in the main text). 
		(\textbf{d}) Knowing the junctions' instantaneous currents (\textbf{b}) and their resistive states (\textbf{c}), we tracked the evolution of the power dissipated by single junctions during the heating plateau (\textbf{a}). While layer $B$ is initially ``cold'', i.e.\ no local hot-spots, layer $A$ has a distribution of hot-spots across the array whose presence becomes more and more dominant the more the system approaches the SC-to-N discontinuous jump; hot-spots propagate between layers as in interdependent percolation, yielding the mutual SN-phase transition reported at equilibrium in the resistive curves (Fig.~\ref{fig:2}\textbf{d}) and mutual phase diagrams (Fig.~\ref{fig:3}\textbf{a}) in the main text. }\label{fig:S2}
	\end{minipage}
\end{figure}

\newpage
\begin{figure}[h!]\vspace*{+3cm}
	\centering
	\sbox\mysavebox{
	\includegraphics[width=0.85\linewidth]{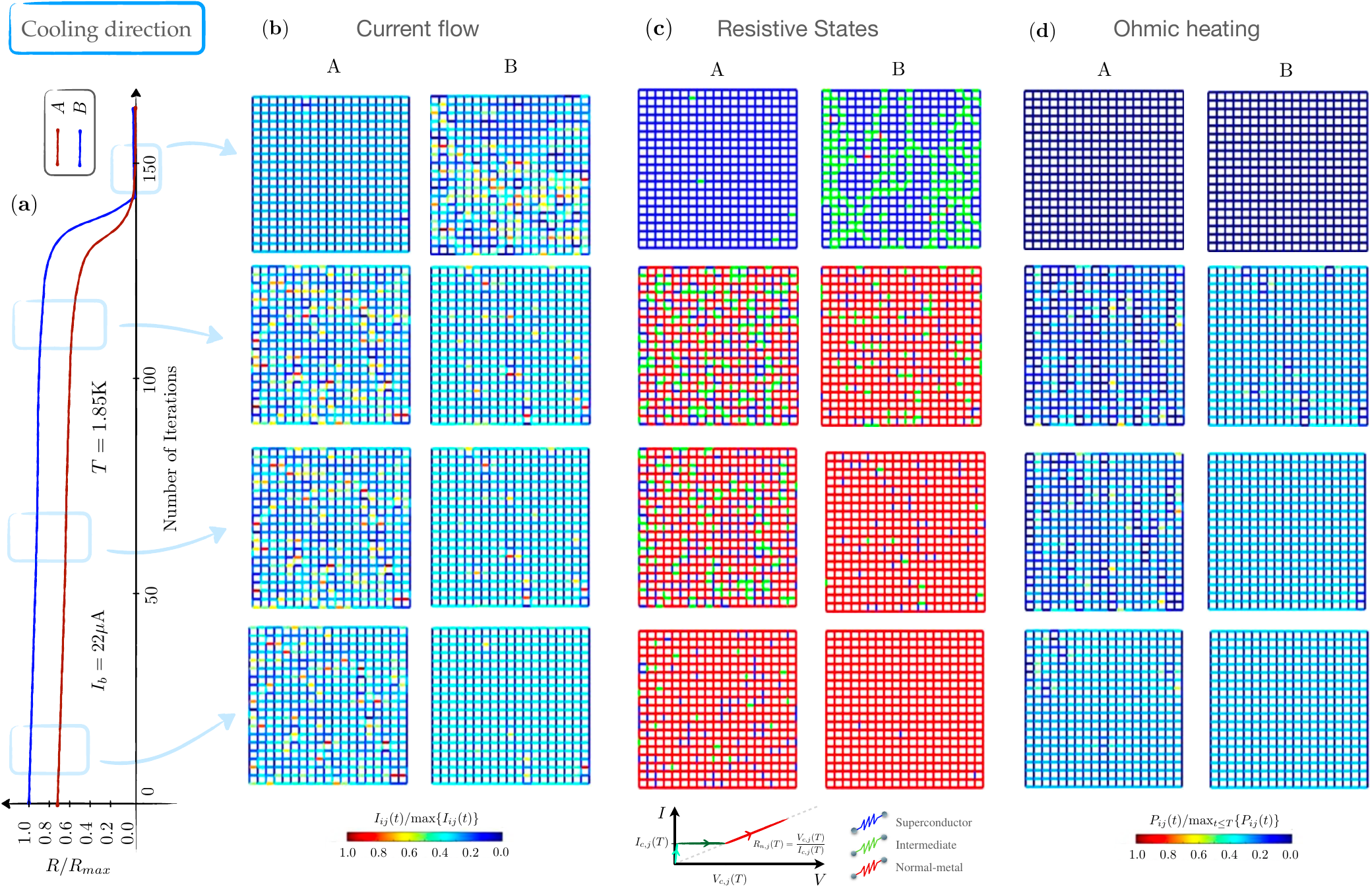}}%
	\usebox\mysavebox
  	\par
  	\begin{minipage}{\wd\mysavebox}\vspace*{+0.3cm}
		\caption{\small (Color online) {\bf Cooling plateau and junctions' state evolution.} 
		(\textbf{a}) Evolution of the marginally stable mutual N-phase of $2$ thermally-interdependent RSJJ networks deep quenched slightly {\em below} the bulk SC-to-N first-order transition threshold, $T_{c,<}=1.88\mathrm{K}$. By contrast with the heating plateau, here the evolution of the metastable phase is faster, though it still undergoes a long-lived metastable stage. 
		(\textbf{b}) By direct inspection of the instantaneous local currents, we find that the current redistribution is less heterogeneous than what found during the heating plateau (Fig.~\ref{fig:S2}\textbf{b}), since the currents search now for paths of minimal resistance. The formation of such paths (see also Supplementary Movies S1 and S2) depends on the number of junctions switching to the SC-state (or to the intermediate state) which tend to be suppressed due to the overheating effect induced by the interdependent couplings. 
		(\textbf{c}) The previous effect yields the formation of many isolated SC-islands in a sea of N-state junctions (see green and blue regions in the red sea of N-state junctions) whose density grows until their sudden merging, resulting in the percolation of a SC-path. Once the latter is formed, the system evolves exponentially rapidly to the mutual SC-phase (for more details, see Fig.~\ref{fig:S4}\textbf{a},\textbf{b}). 
		(\textbf{d}) At the early stages of the mutual N-state, both layers are ``hot'', i.e.\ each junction of the two arrays is a local hot-spot for the superposed one. During the evolution, the formation of local SC-islands produces cooler areas (deep blue paths) which increase the susceptibility of superposed junctions to switch to the SC-state. The propagation of cooler areas---i.e.\ regions whose local temperature corresponds to the one of the cryostat---continues until one or more SC-islands merge together resulting into the abrupt and simultaneous emergence of global phase coherence in the two arrays.
		}\label{fig:S3}
	\end{minipage}
\end{figure}

\newpage
\begin{figure}[h!]\vspace*{+3cm}
	\centering
	\sbox\mysavebox{
	\includegraphics[width=0.85\linewidth]{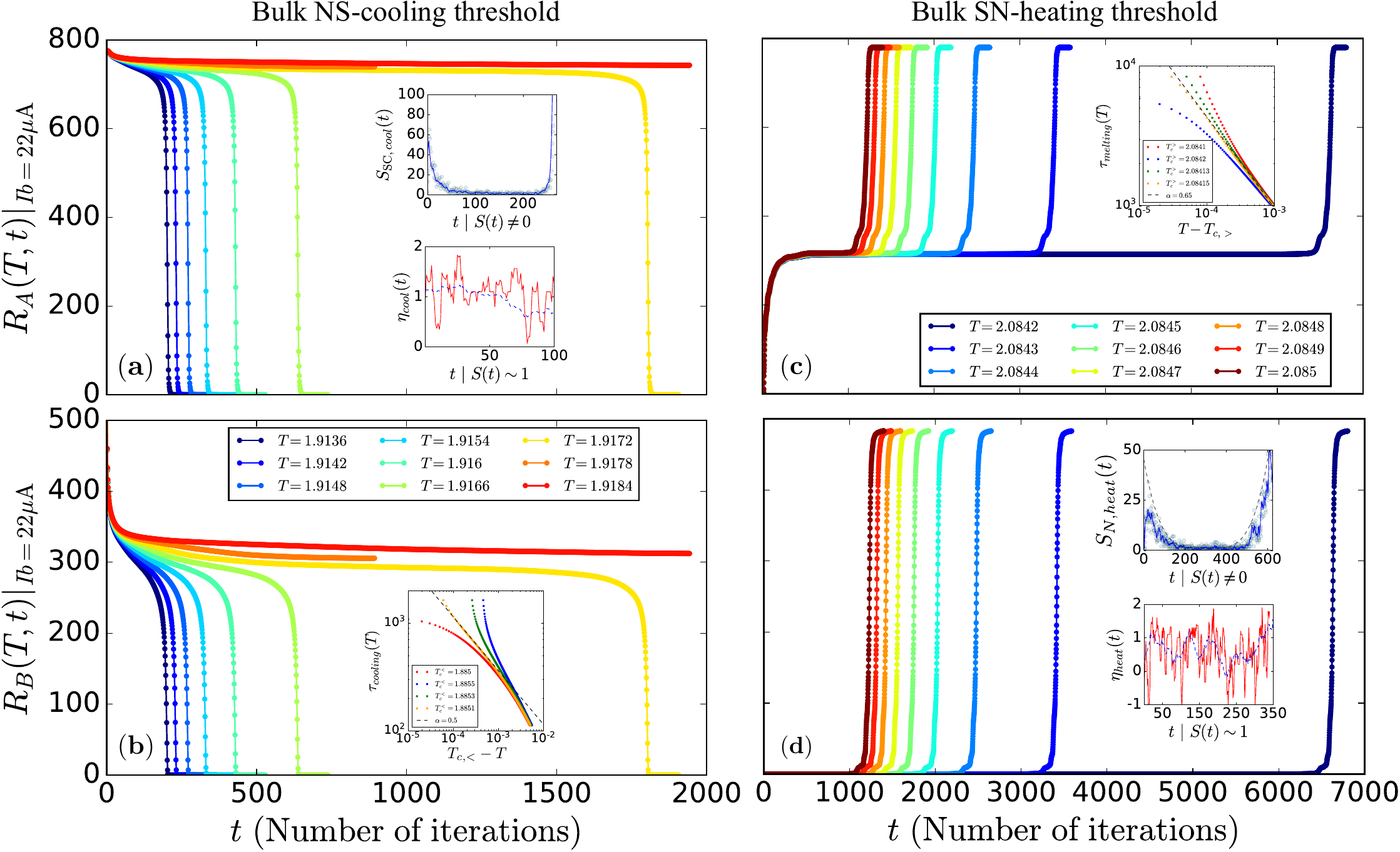}}%
	\usebox\mysavebox
  	\par
  	\begin{minipage}{\wd\mysavebox}\vspace*{+0.3cm}
		\caption{\small (Color online) {\bf Metastable lifetime of the mutual resistive phases}. 
		(\textbf{a}/\textbf{b}) Lifetime of the metastable mutual N-phase in thermally-interdependent RSJJs networks when deep quenching the system at temperatures slightly below the bulk N-to-SC cooling threshold, $T_{c,<}=1.88\mathrm{K}$ in (\textbf{a}) the more disordered layer and (\textbf{b}) the less disordered layer. Different colors describe the behavior observed when departing more and more from $T_{c,>}$ (see the color legend in \textbf{b})). 
		(Inset, \textbf{a}, top) Evolution of the number of SC-state in an array with $L=50$: notice the parabolic behavior (a classical fingerprint of cascading like effects). 
		(Inset, \textbf{a}, bottom) Branching order parameter at the cooling plateau, defined as $\eta_{cool}(t)=\#_{SC}(t)/\#_{SC}(t-1)$, estimating the number of bonds switching to the SC-state at each iteration; notice that, along the plateau, $\eta_{cool}\sim1$, i.e.\ the branching is critical during the metastable evolution. 
		(Inset, \textbf{b}) Scaling of the metastable lifetime obtained by selecting the number of iterations during which the bulk resistance in layer $A$ is nearly constant (similar results are obtained when considering only layer $B$). The lifetime diverges at the first-order transition threshold in MF-fashion, i.e.\ $\tau_{cool}\sim(T_{c,>}-T)^{-\alpha}$ with $\alpha=1/2$~\cite{buldyrev-nature2010, baxter2015critical} \!\!. 
		(\textbf{c}/\textbf{d}) Lifetime of the metastable mutual SC-phase when deep quenched slightly above the bulk SC-to-N heating threshold, $T_{c,>}=2.08\mathrm{K}$ in (\textbf{c}) layer $A$ and (\textbf{d}) layer $B$. 
		(Inset, \textbf{c})Similarly to the cooling plateau, we report also here the emergence of a long-lived metastable state with nearly constant resistance, whose lifetime diverges when approaching $T_{c,>}$ as $\tau_{heat}\sim(T-T_{c,>})^{-\alpha}$ with $\alpha=0.65$. The higher exponent w.r.t.\ the cooling case (for the same system sizes and parameters settings) indicates a slower growth of the SC-nuclei which can be understood as due to the pinning of the N-islands shown in the snapshots in Fig.~\ref{fig:S2}\textbf{c}. 
		(Inset, \textbf{d}, top) Evolution of the number of N-junctions during the heating plateau stage, showing the cascading process of normal-metal switches due to the propagation of heat between layers carried by the hot-spots (red bonds in Fig.~\ref{fig:S2}\textbf{d}). 
		(Inset, \textbf{d}, bottom) Branching OP for the evolution of the N-junctions in Fig.~\ref{fig:S4}\textbf{d} (top inset), showing also here a critical mean branching behavior during the metastable lifetime of the mutual SC-phase.  
		}\label{fig:S4}
	\end{minipage}
\end{figure}

\begin{figure}[h!]
	\centering
	\includegraphics[width=0.75\linewidth]{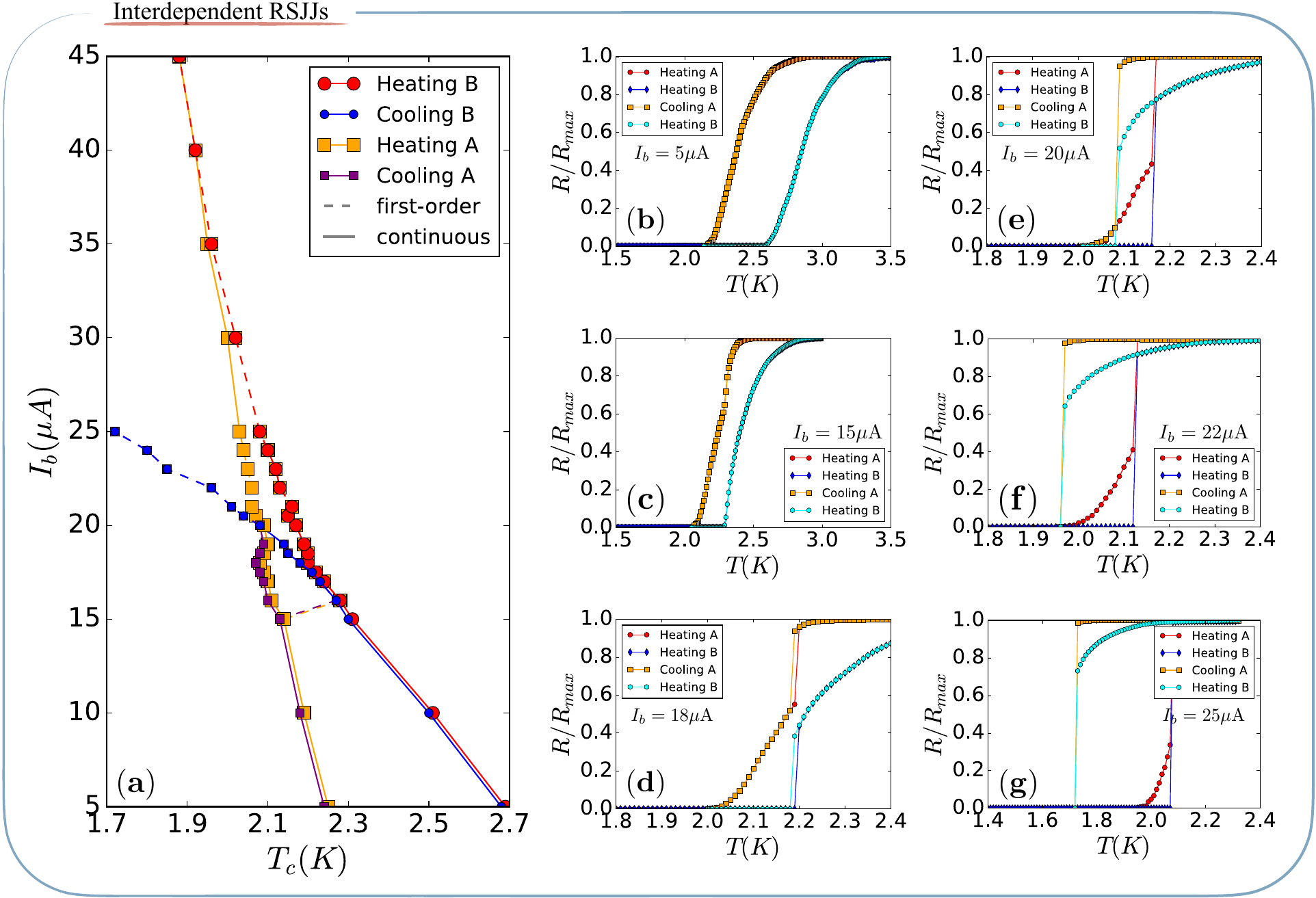}\vspace*{+0.1cm}
	\sbox\mysavebox{
	\includegraphics[width=0.75\linewidth]{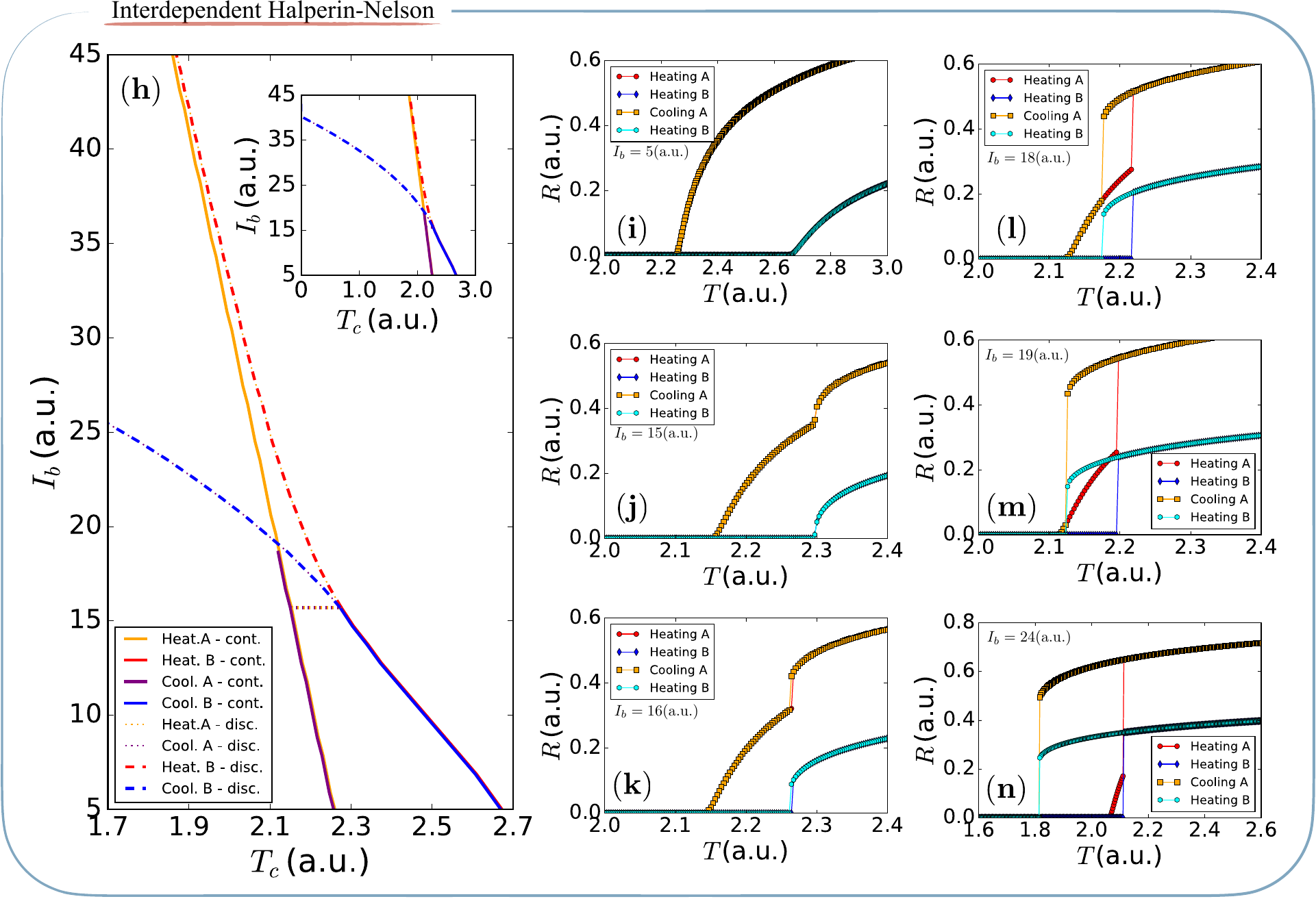}}%
	\usebox\mysavebox
  	\par
  	\begin{minipage}{\wd\mysavebox}\vspace*{+0.2cm}
		\caption{\footnotesize (Color online) {\bf Numerical vs.\ analytical mutual phase diagram.} 
		(\textbf{a}) Mutual phase diagram obtained by solving numerically the thermally-interdependent Kirchhoff equations, Eqs.~\eqref{eq:M1}--\eqref{eq:M3}, for $2$ RSJJ networks coupled via global thermal interactions (see text); full (dashed) lines denote continuous (discontinuous) transitions. 
		(\textbf{b})--(\textbf{g}) Mutual resistive behaviors and transition stages obtained by considering specific values of the driving currents $I_b$ injected in the two arrays; notice the thermal locking of the bulk critical temperatures at sufficiently large currents. 
		(\textbf{h}) Mutual phase diagram solving the steady state of the interdependent Halperin-Nelson resistive equations, Eq.~\eqref{eq:4}, whose material parameters (see text) have been extrapolated by best-fitting the numerical curves of single layers in Fig.~\ref{fig:S5}\textbf{b}--\textbf{g}. Physical quantities ($R$, $T$, $I_b$, etc.) are measured in arbitrary units (a.u.). 
		(Inset, \textbf{h}) Zooming-out the phase diagram: notice the vanishing of the cooling N-to-SC transition at the driving current $I_{b}'\simeq 40(\mathrm{a.u.})$. Above this current, a system prepared in the mutual N-phase remains trapped in the latter without undergoing a N-to-SC transition (see Fig.~\ref{fig:3}\textbf{g}). 
		(\textbf{i})--(\textbf{n}) Mutual resistive behaviors characterizing the phase diagram in \textbf{h}; notice the striking agreement with the numerical (Fig.~\ref{fig:S5}\textbf{a}--\textbf{g}) and the experimental results (Fig.~\ref{fig:1}\textbf{f} and Fig.~\ref{fig:3}\textbf{b}) within the accessible range of parameters. See also Fig.~\ref{fig:S7} for a side-by-side comparisons. 
		}
		\label{fig:S5}
	\end{minipage}
\end{figure}

\newpage
\begin{figure}[h!]\vspace*{+3cm}
	\centering
	\sbox\mysavebox{
	\includegraphics[width=0.85\linewidth]{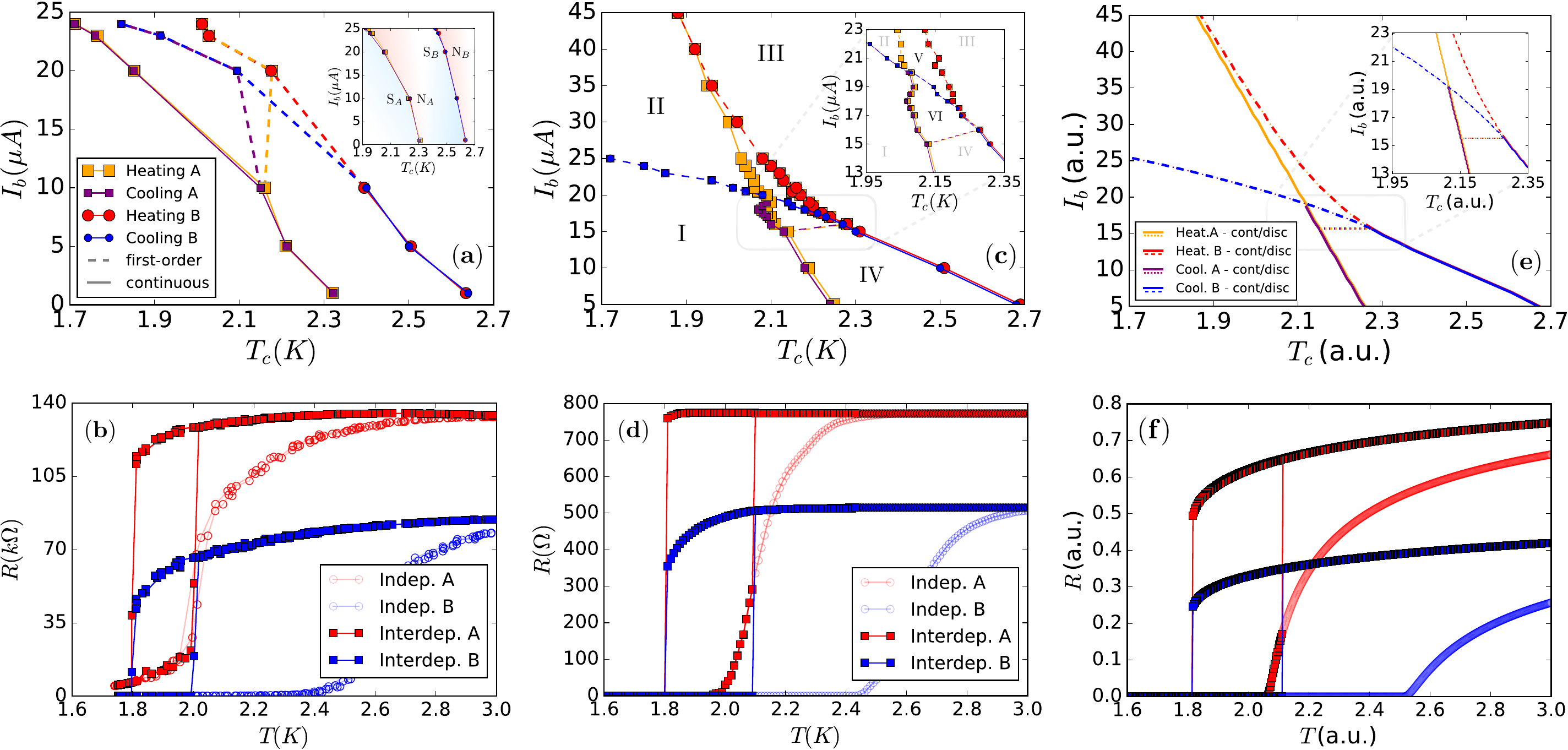}}%
	\usebox\mysavebox
  	\par
  	\begin{minipage}{\wd\mysavebox}\vspace*{+0.3cm}
		\caption{\small (Color online) {\bf Experimental, numerical and analytical results.}
		(\textbf{a}) Experimental phase diagram describing the mutual electronic phases found in our system of $2$ thermally-interdependent superconductors (Fig.~\ref{fig:1}\textbf{f}). For driving currents between $I_b\simeq10\mu\mathrm{A}$ and $I_{b}\simeq 20\mu\mathrm{A}$, both networks undergo discontinuous SN-transitions at thermally-locked critical temperatures (notice the bifurcation of curves from full lines towards dashed lines). 
		(Inset) Collective phases in the disordered networks in \textbf{a} when considered isolated from each other; notice that both layers undergo only continuous SN-phase transitions. 
		(\textbf{b}) Experimental resistive curves describing the mutual transitions reported in the interdependent superconducting network at the driving current $I_b=24\mu\mathrm{A}$. Red (blue) symbols mark the resistive behaviors of the more disordered (less disordered) layer, while empty symbols describe the resistive behaviors when the layers are instead thermally-decoupled (Fig.~\ref{fig:1}\textbf{a},\textbf{c}). Notice that the more disordered array (red symbols) drives the mutual SC-to-N transition after having entered already a partial N-phase at temperatures $T<1.7\mathrm{K}$; on the other hand, when cooling down the system from the mutual N-phase, the less disordered array (blue symbols) drives the system towards a global phase coherent state at the bulk cooling threshold $T_{c,>}\simeq1.8\mathrm{K}$. 
		(\textbf{c}) Numerical phase diagram obtained by solving the thermally-interdependent Kirchhoff equations, Eqs.~\eqref{eq:M1}--\eqref{eq:M3}, describing the electronic state of $2$ thermally-coupled RSJJ networks whose parameters match those of the experiment (see Methods, M4 and M5, as well as Eqs.~\eqref{eq:1} in the main text and discussions therein). 
		(Inset) Close-up into the mutual phase space, highlighting the path from independent towards full thermal-locking in the layers' bulk critical temperatures (compare with Fig.~\ref{fig:S7}\textbf{a}). For a definition of the mutual electronic phases I--VI, we refer the reader to the main text. 
		(\textbf{d}) Alike \textbf{b}) we report here the resistive curves in our interdependent network of RSJJs at $I_b=24\mu\mathrm{A}$; notice the two-stage transition governing the route towards the N-phase (see also Fig.~\ref{fig:S5}\textbf{b}--\textbf{g}) in the more disordered array (red symbols). 
		(\textbf{e}) Analytical phase diagram obtained by solving the interdependent Halperin-Nelson equations for the bulk sheet-resistance, Eq.~\eqref{eq:4}. Alike what described in Fig.~\ref{fig:S5}\textbf{h}, the material parameters have been selected after best fitting the thermally-decoupled resistive behaviors to the numerical curves reported in the numerical solutions to of the independent RSJJ networks; the thermal coupling is best-fitted to track the trend of heat-dissipation for increasing values of the driving current (see Eq.~\eqref{eq:4} in the main text and details therein). 
		(Inset) Close-up in the mutual phase space where we report the onset of thermal-locking of the bulk critical temperatures; notice the striking agreement with the numerical and experimental results (Fig.~\ref{fig:S7}\textbf{c}, Inset). 
		(\textbf{f}) Interdependent Halperin-Nelson resistive curves obtained at the driving current $I_b=23 (\mathrm{a.u.})$, showing the general emergence of a two-stage transition in the path from the full SC-state to the full N-state in the more disordered array (red symbols) and the drive of the less disordered array in the path from the mutual N-phase to the full SC-phase. Curves striking agree qualitatively and nearly quantitatively with the numerical and experimental results. 
		}
		\label{fig:S7}
	\end{minipage}
\end{figure}

\end{document}